\documentclass{aa}  

\usepackage{graphicx}
\usepackage{txfonts}
\usepackage{cprotect}
\usepackage[table]{xcolor}
\usepackage{orcidlink}
\usepackage{multirow}
\newcommand{\mjup}{M$_{\rm Jup}$\,}

\newcommand{\straklip}{\textit{Stra}KLIP}
\newcommand{\publicwifi}{\textit{PUBLIC-WiFi}}
\newcommand{\orcid}[1]{\orcidlink{#1}}

\begin{document}

   \title{Multiplicity of young brown dwarfs and isolated planetary mass objects in Taurus and Upper Scorpius}

   \author{H. Bouy\orcid{0000-0002-7084-487X}\inst{1, 2}
          \and
          G. Duch\^ene\orcid{0000-0002-5092-6464}\inst{3,4} 
        \and G. Strampelli\orcid{0000-0002-1652-420X}\inst{5} 
        \and J. Aguilar\orcid{0000-0003-3184-0873}\inst{5}        
          \and J. Olivares\orcid{0000-0003-0316-2956}\inst{6}   
          \and A. Palau \orcid{0000-0002-9569-9234}\inst{7}
          \and D. Barrado\orcid{0000-0002-5971-9242}\inst{8}
          \and S.~N. Raymond\orcid{0000-0001-8974-0758}\inst{1}
          \and N. Huélamo\orcid{0000-0002-2711-8143}\inst{8}
          \and M. Tamura\orcid{0000-0002-6510-0681} \inst{9,10,11}          
          \and E. Bertin\orcid{0000-0002-3602-3664} \inst{12}
          \and W. Brandner\orcid{0000-0001-6023-4974}\inst{13}
          \and J.-C. Cuillandre\orcid{0000-0002-3263-8645} \inst{12}
          \and P.~A.~B. Galli\orcid{0000-0003-2271-9297}\inst{14}                  
          \and N. Miret-Roig \orcid{0000-0001-5292-0421}\inst{15,16,17} 
          }

   \institute{Laboratoire d'astrophysique de Bordeaux, Univ. Bordeaux, CNRS, B18N, allée Geoffroy Saint-Hilaire, 33615 Pessac, France.\\
              \email{herve.bouy@u-bordeaux.fr}
         \and
         Institut universitaire de France (IUF), 1 rue Descartes, 75231 Paris CEDEX 05
         \and Univ. Grenoble Alpes, CNRS, IPAG, 38000 Grenoble, France
         \and Astronomy Department, University of California Berkeley, Berkeley CA 94720-3411, USA
         \and 
         Space Telescope Science Institute, 3700 San Martin Dr., Baltimore, MD 21218, USA
        \and
         Departamento de Inteligencia Artificial, Universidad Nacional de Educación a Distancia (UNED), c/Juan del Rosal 16, E-28040, Madrid, Spain
         \and
         Universidad Nacional Aut\'onoma de M\'exico, Instituto de Radioastronom\'ia y Astrof\'isica, Antigua Carretera a P\'atzcuaro 8701, Ex-Hda. San Jos\'e de la Huerta, 58089, Morelia, Michoac\'an, M\'exico
        \and 
         Centro de Astrobiología (CAB), CSIC-INTA, ESAC Campus, Camino bajo del Castillo s/n, E-28692 Villanueva de la Ca\~nada, Madrid, Spain 
        \and
         Department of Astronomy, Graduate School of Science, The University of Tokyo, Tokyo, Japan
         \and 
         National Astronomical Observatory of Japan,  Tokyo, Japan   
         \and
         Astrobiology Center, Tokyo, Japan
         \and
         Universit\'e Paris-Saclay, Universit\'e Paris Cit\'e, CEA, CNRS, AIM, 91191, Gif-sur-Yvette, France 
         \and 
         Max-Planck-Institut für Astronomie, Königstuhl 17, 69117 Heidelberg, Germany
        \and
        Instituto de Astronomia, Geofísica e Ciências Atmosféricas, Universidade de São Paulo, Rua do Matão, 1226, Cidade Universitária, 05508-090 São Paulo-SP, Brazil        
         \and 
        Departament de Física Quàntica i Astrofísica (FQA), Universitat de Barcelona (UB), Martí i Franquès, 1, 08028 Barcelona, Spain
        \and
        Institut de Ciències del Cosmos (ICCUB), Universitat de Barcelona (UB), Martí i Franquès, 1, 08028 Barcelona, Spain 
        \and
        Institut d'Estudis Espacials de Catalunya (IEEC), Edifici RDIT, Campus UPC, 08860 Castelldefels (Barcelona), Spain
             }

   \date{Received ; accepted }

% \abstract{}{}{}{}{} 
% 5 {} token are mandatory
 
  \abstract
   {Free-floating planetary mass objects--worlds that roam interstellar space untethered to a parent star--challenge conventional notions of planetary formation and migration, but also of star and brown dwarf formation. }
   {We focus on the multiplicity among free-floating planets. By virtue of their low binding energy (compared to other objects formed in these environments), these low-mass substellar binaries represent a most sensitive probe of the mechanisms at play during the star formation process.}
   {We use the {\it Hubble Space Telescope} and its Wide Field Camera 3 and the {\it Very Large Telescope} and its ERIS adaptive optics facility to search for visual companions among a sample of 77 objects members of the Upper Scorpius and Taurus young nearby associations with estimated masses in the range between approximately 6--66~\mjup.}
   {We report the discovery of one companion candidate around a Taurus member with a separation of 111.9$\pm$0.4~mas, or $\sim$18~au assuming a distance of 160~pc, with an estimated primary mass in the range between 3--6~\mjup and a secondary mass between 2.6--5.2~\mjup, depending on the assumed age. This corresponds to an overall binary fraction of 1.8$^{+2.6}_{-1.3}$\% among low-mass brown dwarfs and free-floating planetary mass objects over the separation range $\ge$7~au.  Despite the limitations of small-number statistics and variations in spatial resolution and sensitivity, our results, combined with previous high-spatial-resolution surveys, suggest a notable difference in the multiplicity properties of objects below $\sim$30--50~\mjup between Upper Sco and Taurus. In Taurus, a binary fraction of $5.6^{+3.2}_{-2.3}$\% is found for objects with masses below 30\mjup, and of $7.8^{+3.0}_{-2.4}$\% for objects with masses below 50\mjup, whereas no binary were found among 80 objects over the matching luminosity range in Upper Sco, corresponding to an upper limit of $\le$1.2\%.}
    { This difference may point to intrinsically distinct formation conditions, with warmer parental molecular clouds originally present in Upper Sco potentially inhibiting fragmentation into the lowest-mass brown dwarfs and free-floating planets compared to cooler environments such as Taurus.}
    

   \maketitle
%
%-------------------------------------------------------------------

\section{Introduction}
Free-floating planetary mass objects (FFPs), also called isolated planetary mass-objects (IPMOs), are planetary-mass objects that do not orbit a star, but roam the galaxy isolated. They are some of the most challenging astrophysical objects to study: incapable of sustaining nuclear fusion, they are intrinsically extremely faint and steadily fade over time to become very difficult to detect directly. Young FFPs with super-Jovian masses in star forming regions had been discovered using early direct imaging \citep[e.g.][]{Tamura1998, Lucas2000, Zapatero2000}. In contrast, less-massive FFPs had been discovered using indirect methods: gravitational micro-lensing surveys have proven particularly successful to detect them down to a few Earth masses \citep[e.g.][ and references therein]{Mroz2017,Mroz2020,Ryu2021,Sumi2023}. The future {\it Roman} Space Telescope will be sensitive to FFP down to masses as low as 0.1~M$_{\oplus}$ and is predicted to harvest tens of such objects during the course of its micro-lensing survey \citep{Johnson2020}. The advent of large and very deep ground-based surveys led to the discovery of several hundreds FFP candidates in nearby star forming regions \citep[][ and references therein]{PenaRamirez2012, Scholz2012, Muzic2011,  Muzic2015,Bayo2011, Esplin2019, MiretRoig2022} and the field \citep{Delorme2008,Delorme2017}. The WISE survey \citep{Wright2010} and its subsequent extensions led to the discovery of several FFPs in the solar neighborhood \citep{Beichman2013,Schneider2016}. More recently the JWST and Euclid missions are starting to reveal large samples of FFPs thanks to their unprecedented sensitivities in the infrared \citep{Bouy2025, Luhman2024, Martin2024,Langeveld2024,DeFurio2024,DeFurio2025}.

The origin and the formation of FFP remains poorly understood. Even though they represent a very small fraction of the total baryonic mass of our galaxy, they add up to at least 2 to 5\% of the astrophysical objects populating the Milky Way \citep{MiretRoig2022}, making them more numerous than stars with masses greater than 3~M$_{\odot}$. Similar to slightly more massive brown dwarfs, several scenarios are considered to explain their existence:

\begin{itemize}
\item[--] a scaled-down version of star formation through gas cloud direct collapse and turbulent fragmentation \citep{Padoan2004, Hennebelle2008};
\item[--] formation within a proto-planetary disc, either like gas-giant planets through core accretion (Pollack et al., 1996) or like companions through gravitational fragmentation of massive extended discs \citep{Boss1998}, followed by ejection by dynamical scattering between planets in both cases \citep{Veras2012};
\item[--] as aborted stellar embryos ejected from a stellar nursery before the hydrostatic cores could build up enough mass to become a star \citep{Reipurth2001};
\item[--] through the photo-erosion of a pre-stellar core by stellar winds from a nearby OB star before it can accrete enough mass to become a star \citep{Whitworth2004}
\item[--] within dense filaments produced by very close encounters between gaseous protoplanetary disks \citep{Fu2025}.
\end{itemize}

While recent direct observational evidence confirms that these different processes are at work \citep[][ and references therein]{MiretRoig2022,Palau2024,Bouy2009}, we still do not understand their relative contributions to the overall FFP population (e.g., photo-erosion can only occur in the direct vicinity of relatively scarce OB stars) and how the environment affects their outcome. 

Historically, multiple systems have been commonly used to to test and constrain the
theories and simulations of formation and evolution. Multiple systems are indeed key diagnostics of star formation in general and of BDs and FFPs in particular. According to theory and simulations, multiple systems of isolated planetary mass objects are difficult to form and easy to break \citep{Bate2012}. The main formation scenarios for FFP  allow only few ($<$15\%) and tight ($<$10~au) binaries to form or survive. In the ejection scenario (from a disc, a planetary system or a cluster), most primordial binaries would indeed be disrupted by dynamical interactions and only systems with the highest binding energy survive, resulting in a strong preference for tighter separations and higher mass ratios. In the core-collapse scenario, the multiplicity properties are expected to follow the trends seen for more massive objects, with a binary frequency decreasing with decreasing primary mass, a semi-major axis distribution peaking at closer separations and mass ratios shifting towards unity \citep{Duchene2013,Offner2023}.

Based on these considerations, FFP binaries should be rare and tight and hence difficult to detect. Recent observations using the {\it Hubble Space Telescope} did not detect any visual binaries among a sample of 33 field T and Y-dwarfs \citep{Fontanive2023} with estimated masses in the range 11--46~\mjup. However, observations of some young multiple systems with primaries near or within the planetary-mass regime suggest a different scenario. Several binary candidates with primaries close to or below the planetary-mass limit exhibit surprisingly large separations and relatively small mass ratios \citep{Close2007, Fontanive2020, Langeveld2024, Luhman2024b}. This contrasts sharply with theoretical and numerical predictions, as well as with the properties of slightly more massive, older brown dwarf binaries, which have a multiplicity fraction of approximately 10--15\% and a tight separation distribution centered around 3–7~au. \citep{Duchene2013, Fontanive2018, Offner2023}. 

Due to limited sample sizes and the absence of systematic studies, it is unclear whether the discovery of these young, wide free-floating planetary mass binaries reflects a fundamental change in multiplicity properties within the planetary-mass regime or at younger ages, represents rare systems uncharacteristic of the broader FFP binary population, or is merely the result of observational biases favoring large separations. To answer these questions, we investigate the multiplicity among a sample of 77 young FFP and BD members of the Taurus and Upper Scorpius (USco) associations using the {\it Hubble Space Telescope} (HST) and the {\it Very Large Telescope} (VLT). 

\section{Targets}
\subsection{HST targets}
\begin{figure}
\centering
\includegraphics[width =0.49\textwidth]{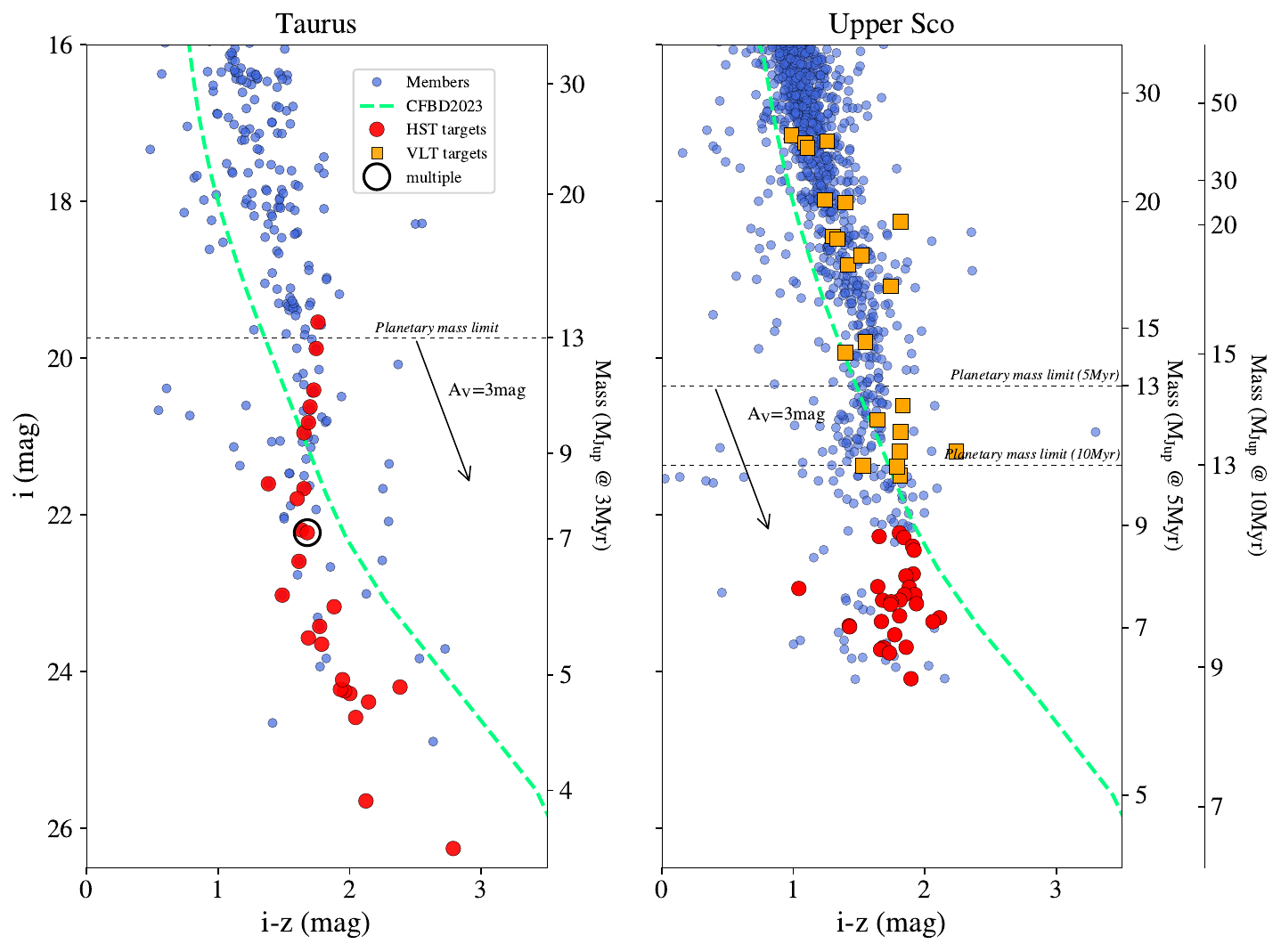}
\caption{ ({\it i, i-z}) diagram of Taurus members (left) identified by \citet{Esplin2019} and USco members (right) from \citet{MiretRoig2022}, represented as blue dots. The HST targets are over-plotted as red dots and the VLT targets are orange squares. The \citet{Chabrier2023} isochrones at 3~Myr (Taurus) and 5 and 10~Myr (USco) and 140~pc are represented by  green dashed lines and the corresponding masses are indicated on the right vertical axis. The 5 and 10~Myr isochrones for USco overlap almost perfectly and are represented as one. A reddening vector A$_{\rm V}$=3~mag is also represented, and the planetary mass limit of 13~\mjup is indicated. The new binary candidate identified in Taurus is over-plotted as a black open circle. 
\label{fig:targets_cmd}}
\end{figure}

We selected 30 targets in the USco sample of FFPs from \citet{MiretRoig2022}, and 30 targets in the sample of Taurus brown dwarfs and FFP candidates from \citet{Esplin2019} and Bouy et al. (in prep.). The latter used the same methodology and strategy as \citet{MiretRoig2022} based on proper motions and multi-wavelength photometry in the Taurus molecular clouds to identify high-probability candidate members down to a few Jupiter masses only. Figure~\ref{fig:targets_cmd} shows ({\it i, i-z}) colour-magnitude diagrams of the targets in each region. Because both samples are based on a selection using both proper motions and multi-wavelength photometry and given the relatively large mean proper motions of Taurus and USco members with respect to field objects, the contamination rate by background or foreground sources unrelated to theses star forming regions is expected to be very low \citep[][]{Bouy2022}.

In USco, we randomly selected targets below the planetary mass limit plus 3~mag of extinction, to minimize contamination by reddened more massive objects. The same strategy was initially used for the Taurus sample, but the lack of suitable HST guide stars in the much sparser Taurus region forced us to drop some targets in favor of slightly more massive objects with guide stars available. 

Recent {\it Gaia} results have revealed that the various subgroups within these two associations span a wide range of distances (from 130 to 200~pc, with an average around 140~pc) and ages, but overlap on the plane of the sky and in the proper motion space. The Taurus association includes groups with ages ranging mostly from 1 to 3Myr \citep{Luhman2023}, while USco includes sub-groups with ages between 3--20~Myr, with a median age around 10~Myr \citep{Schmitt2022, MiretRoig2022b,Luhman2022, Zerjal2023, Ratzenbock2023}. Our targets lack sufficient data (specifically parallax and radial velocity measurements) to determine their precise subgroup membership. Therefore, for the purposes of this study and in all subsequent figures, we adopt an average distance of 140~pc for both samples, assume an age of 3~Myr for the Taurus targets and 5--10~Myr for those in Upper Sco.

\begin{figure}
\includegraphics[width =0.5\textwidth]{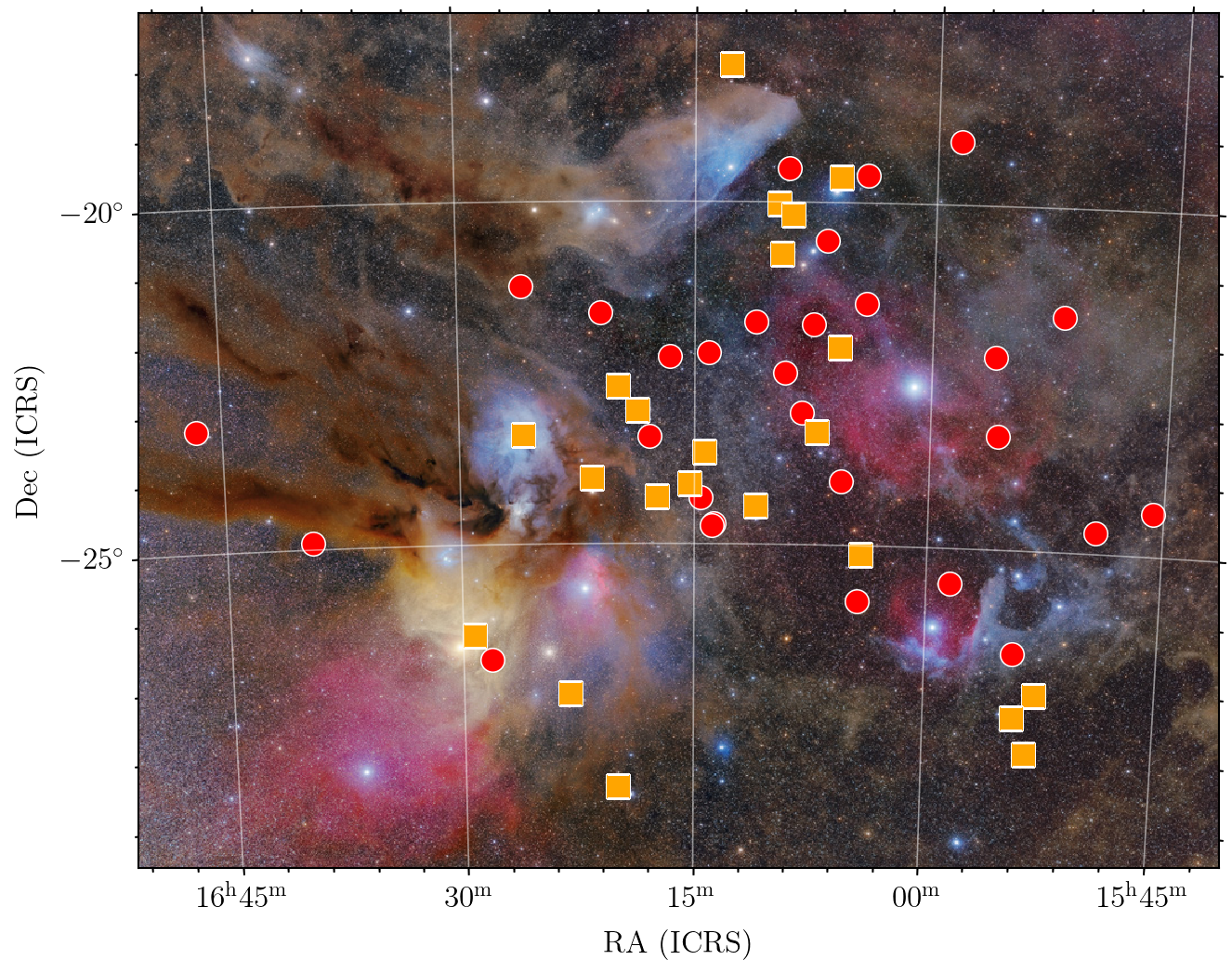}
\caption{Positions of the USco targets on a color photograph showing the clouds and nebulae. HST targets are represented with red dots and VLT targets with orange squares.  Background photograph credit: Mario Cogo. \label{fig:usco_map} }
\end{figure}

\begin{figure}
  \centering 
  \includegraphics[width=0.5\textwidth]{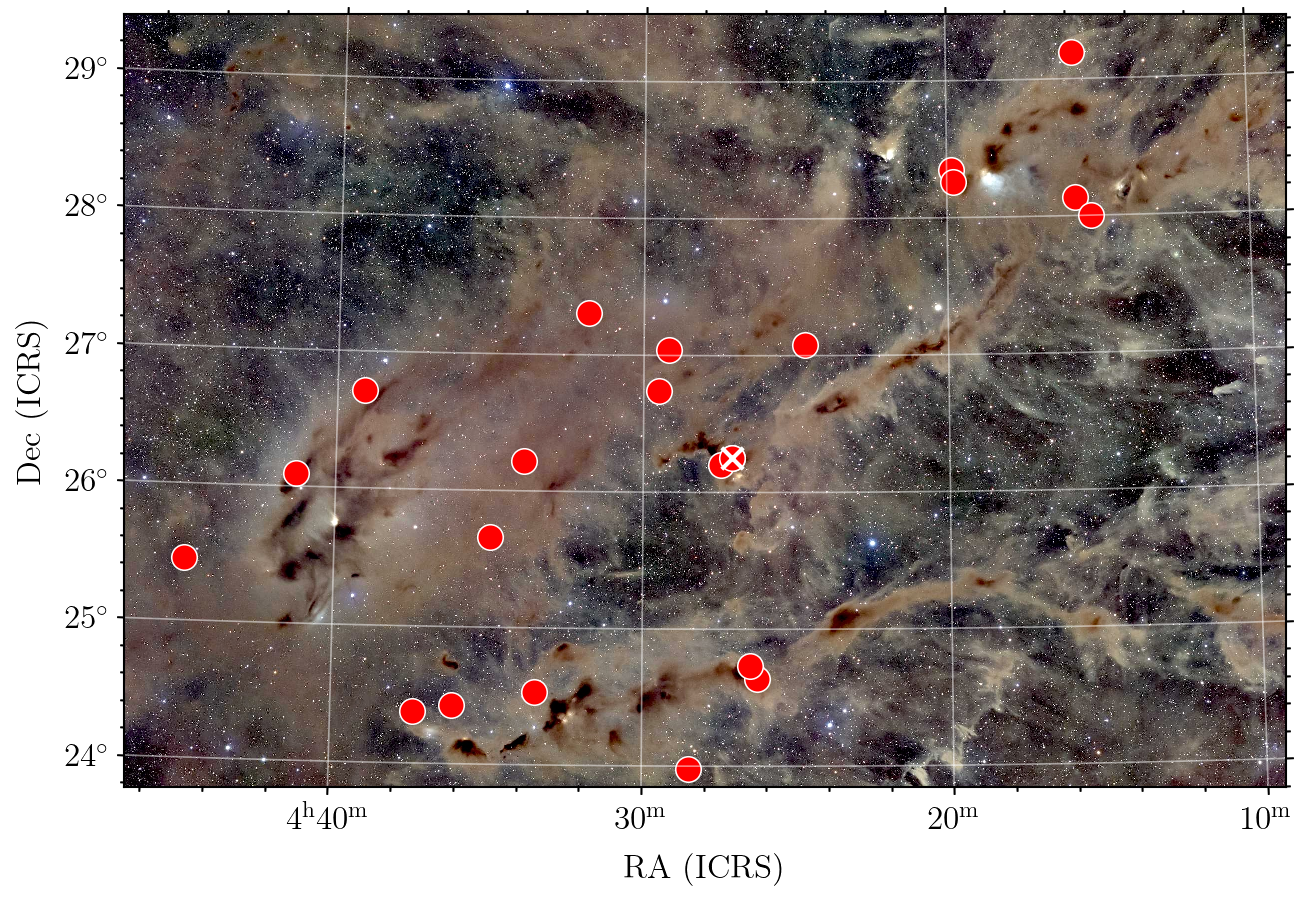} 
  \caption{Positions of the Taurus targets on a color photograph showing the main Taurus molecular clouds. HST targets are represented with red dots and the new binary candidate is indicated with a white cross. Background photograph credit: Chris McGrew} 
  \label{fig:taurus_map} 
\end{figure}

The USco targets thus sample the mass range between 6--66~\mjup and the Taurus sample between 5--33~\mjup, estimated following the method described in \citet{MiretRoig2022} by analyzing the available multi-wavelength photometry with \emph{Sakam} \citep{sakam} and the ages and distances mentioned above. Figures~\ref{fig:usco_map} and \ref{fig:taurus_map} show that the targets are randomly distributed in each association and form a representative sample of the overall associations and the sub-groups composing them.

\subsection{VLT targets}
We selected 60 targets from the USco free-floating planet sample identified by \citet{MiretRoig2022}, distinct from the HST sample described above, based upon the availability of a sufficiently bright reference star for the adaptive optics wavefront sensor within 15\arcsec. The selection was also designed to cover the 12--30~\mjup\ mass range, complementing the HST sample mentioned earlier and prior brown dwarf multiplicity studies in USco by \citet{Bouy2006, Kraus2006, Biller2011, Kraus2012}.

%--------------------------------------------------------------------
\section{Observations and data reduction}

\subsection{HST WFC3/UVIS}
Each target was observed with the UVIS channel of the WFC3 in the F814W and F850LP filters (program GO 17167, P.I. Bouy). A series of four 178~s exposures per filter was acquired, utilizing the maximum time available within a single HST orbit. Three visits failed because the Fine Guidance Sensors lost lock on the guide stars, and the corresponding data is useless. 
Two faint targets were not or barely detected and are also discarded in the rest of the analysis. The final observed sample thus adds up to 30 targets in USco and 25 in Taurus.      

We retrieved the pipeline processed data from the MAST archive, and used the FLC products corresponding to the calibrated individual exposure including charge transfer efficiency (CTE) correction, and the DRC products corresponding to the drizzled FLC individual images  corrected for geometric distortion. We evaluated the peak signal-to-noise (S/N) of each object by quadratically combining Poisson noise based on the pixel counts and the background noise, estimated as the standard deviation in a surrounding annulus. Most targets are detected with $20 \leq \mathrm{S/N} \leq 50$ (median values of 28 and 26 in the F814W and F850LP filters, respectively), with a few outliers at both ends of the brightness distribution of our targets. 

Because all objects were observed with the same observing sequence, we expect to achieve a similar sensitivity for wide companions, where background noise dominates the noise budget. To evaluate this, we performed injection-and-recovery of companions at wide separations ($\gtrsim0\farcs2$) spanning broad ranges of magnitudes and random position angles around the targets. We test the detectability of companions based on a 5$\sigma$ threshold after least squares subtraction of a single star from these mock-up binaries. From these, we compute the companion magnitude for which we achieve a 50\% probability detectability (as a function of position angle). We find that the detection limit for wide companions is at 26.5$\pm$0.4\,mag and 26.3$\pm$0.3\,mag in the F814W and F850LP filters, respectively, corresponding to a fraction of a Jupiter mass at the ages and distances of Taurus and USco.

\subsection{VLT ERIS/NIX}
Observations with the {\it Enhanced Resolution Imager and Spectrograph} (ERIS) adaptive optics (AO) instrument and its NIX imager \citep{ERIS} were conducted at the VLT in queue mode for 25 of the 60 targets between April and December 2023 as part of program 111.24H3 (P.I. Bouy). However, due to insufficient data quality in 3 cases, the final dataset comprised 22 objects. Table~\ref{tab:vlt-targets} gives a summary of their properties as reported in \citet{MiretRoig2022}. Figure~\ref{fig:usco_map} shows their positions in the association, and Figure~\ref{fig:targets_cmd} shows their positions in a ({\it i, i-z}) color-magnitude diagrams. The targets were too faint for the wavefront sensor and nearby ($<$15\arcsec) bright (Gaia RP$<$16.5~mag) stars were used as reference for the adaptive optics. NIX has a pixel scale of 13~mas, and the observations were conducted in the J band, which provides an optimal balance between sensitivity to the cool, red targets and spatial resolution. Each target was observed with 5 DIT of 5~s exposures at 9 jittered positions, resulting in a total on-source exposure time of 225~s per target. Raw images and associated calibration frames were retrieved from the ESO archive, and the data was processed using the official ESO ERIS pipeline (v1.7.0). Table~\ref{tab:vlt-targets} reports the Strehl ratios of the observations, ranging between 2 and 32\%. 

%--------------------------------------------------------------------
\section{Identification of companions}

\subsection{HST images}
\begin{figure}
\centering
\includegraphics[width =0.49\textwidth]{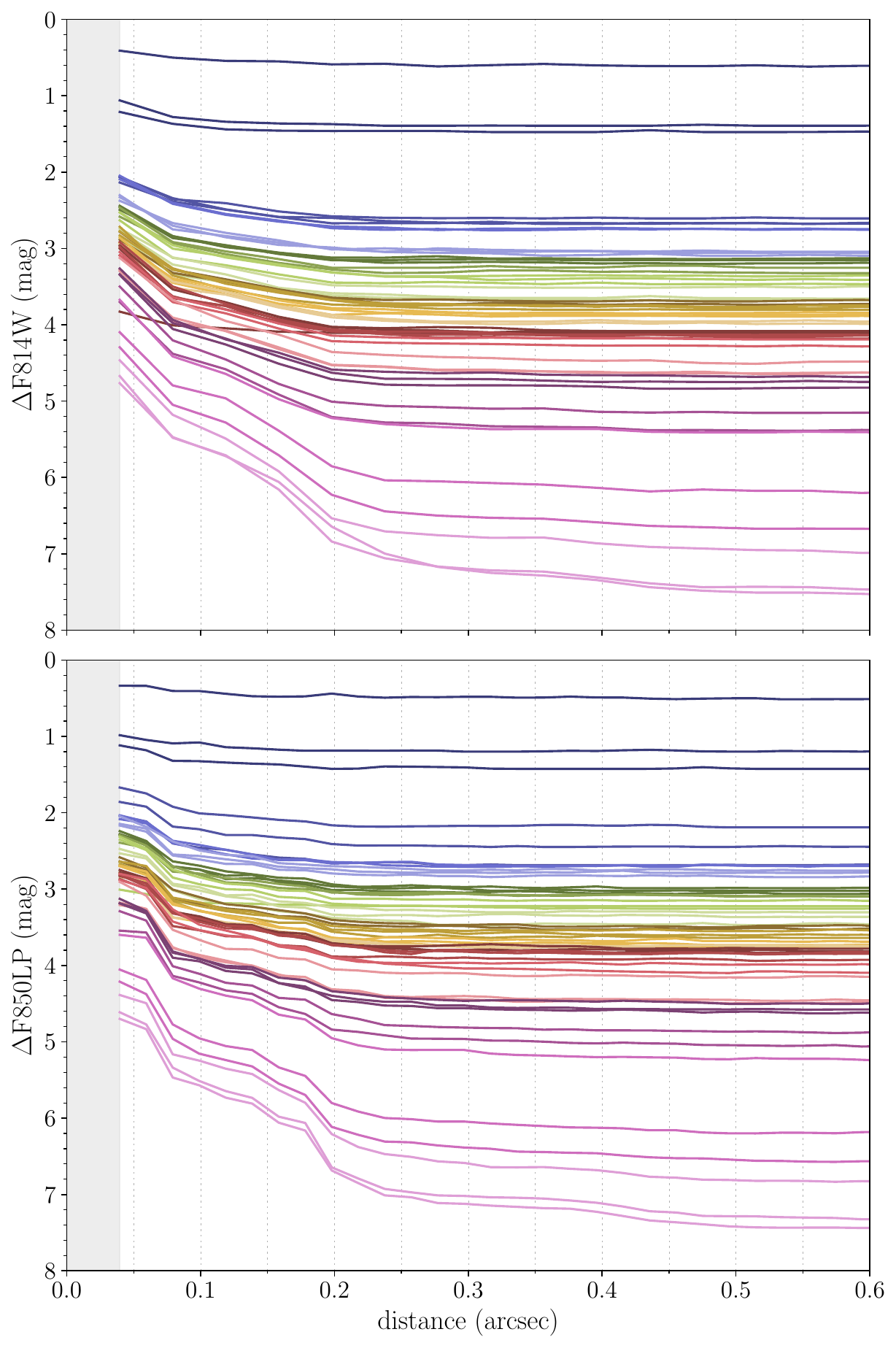}
\caption{Detection limits in the HST F814W (top) and F850LP (bottom) images. }
\label{fig:contrast_hst}
\end{figure}

Figure~\ref{fig:contrast_hst} shows the detection limits determined by computing the average 3$\sigma$ rms noise along the radial profile of each object in the two filters. It shows that, in most cases, our observations are sensitive enough to detect companions with a flux ratio of $\sim$0.5 ($\Delta$mag=0.75~mag) at separations as small as 0\farcs05 (or approximately 7~au at the distance of USco and Taurus). 

We initially visually examined all images to identify easily detectable companions, resulting in the discovery of three visual companions within 1\arcsec\ of their primaries. 

\begin{figure}
\centering
\includegraphics[width =0.48\textwidth]{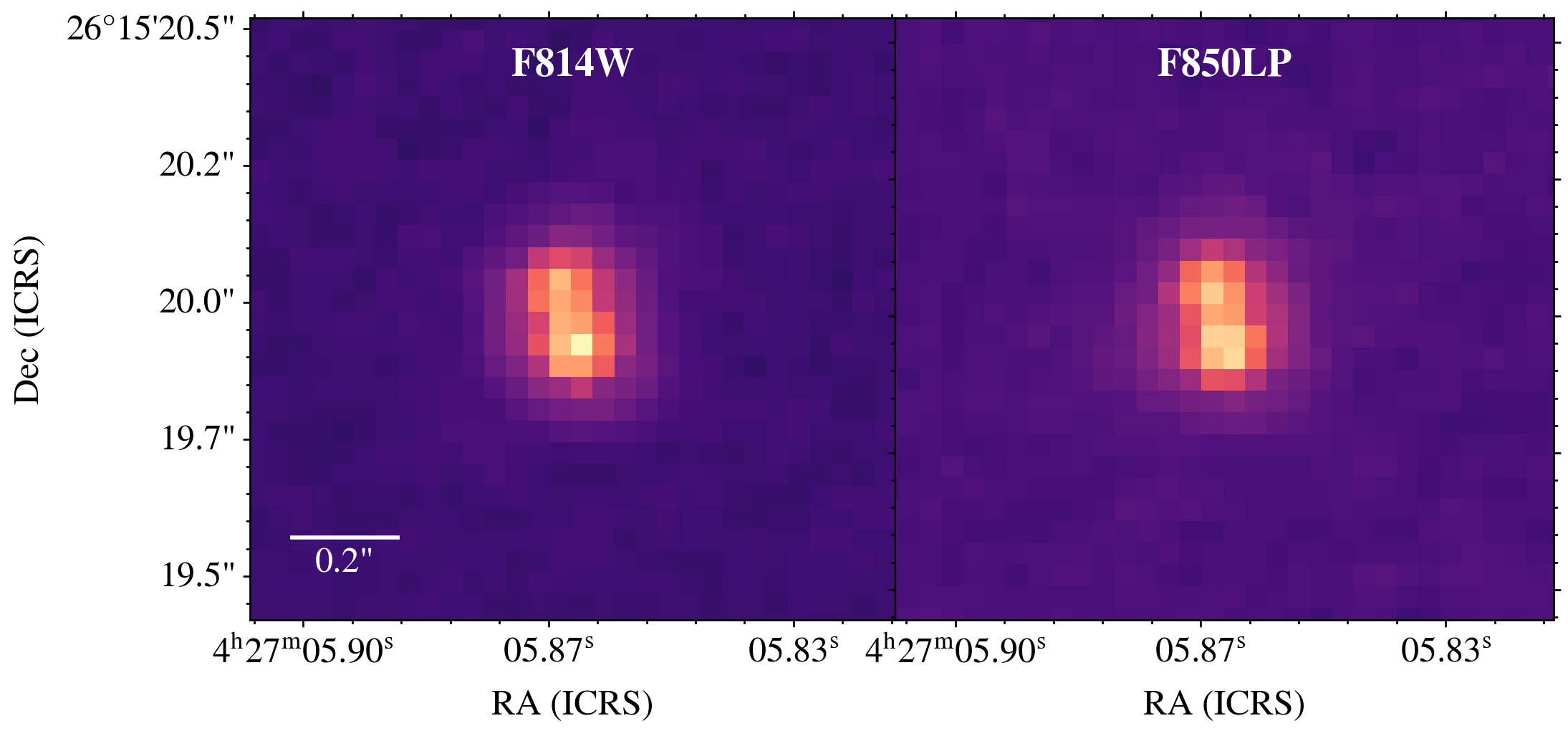}\\
\caption{HST F814W (left) and F850LP (right) images of the companion identified around J042705.86+261520.3  through direct inspection. A square root stretch is used. North is up and east is left. \label{fig:new_binary}}
\end{figure}

J042705.86+261520.3 was confirmed as a binary candidate with a separation of 111.9$\pm$0.4mas, as illustrated in Fig.~\ref{fig:new_binary}. Table~\ref{tab:binaries} presents the relative astrometry and flux ratio measured in the F814W and F850LP images, while Table~\ref{tab:binary_phot} provides the individual component photometry in the two HST filters. J042705.86+261520.3 lies in close proximity, in projection, to the filamentary structure containing clumps B211, B231, B216, B217, and B218, which is connected to LDN1495 and corresponds to Group 8 in \citet{Galli2019}, located at an average distance of 160pc. If this distance is assumed, the measured angular separation translates to a physical separation of approximately 18~au.

We estimated component masses by linearly interpolating the predicted values from the \citet{Chabrier2023} models for ages of 1Myr and 3Myr, assuming a distance of 160pc. The resulting mass range for each component is listed in Table\ref{tab:binary_phot}. These values should be considered lower limits, as extinction could increase them. The 3D extinction map of \citet{Green2019} reports an integrated line-of-sight extinction of $A_V\sim$1.1~mag at the position of J042705.86+261520.3. Assuming a conservative visual extinction of $A_V = 3$~mag would yield a primary mass range of 5--10~\mjup and a secondary mass range of 4.5--9~\mjup.  In any case, the mass ratio is relatively high, approximately 0.9.

\begin{table}
\caption{ J042705.86+261520.3AB properties \label{tab:binaries}}
\small
\centering
\begin{tabular}{rrccc}
\hline
\hline
Parameter & Filter & Value & Uncertainty  & Weighted Average                   \\
\hline
 \multirow{ 2}{*}{$\delta$ (mas)}  & F814W & 112.5 & 0.4 & \multirow{ 2}{*}{111.9$\pm$0.4} \\
 & F850LP  & 111.3 & 0.4 & \\
\hline
 \multirow{ 2}{*}{P.A (\degr)} & F814W   & 14.8 & 0.1 & \multirow{ 2}{*}{13.6$\pm$0.9} \\
 & F850LP     & 12.4 & 0.1 & \\
\hline
 \multirow{ 2}{*}{flux ratio} & F814W    & 0.702 & 0.005 &  \\
 & F850LP      & 0.761 & 0.003 & \\
 \hline
\end{tabular}
\end{table}

\begin{table}
\caption{ J042705.86+261520.3 A and B properties \label{tab:binary_phot}}
\small
\centering
\begin{tabular}{rccccc}
\hline
\hline
Component & F814W & $\sigma_{\rm F814W}$ & F850LP & $\sigma_{\rm F850LP}$ & Mass  \\
        & (mag) & (mag) & (mag) & (mag) & (\mjup) \\
\hline
A & 21.92 & 0.02 & 21.03 & 0.02 & 3--6 \\
B &  22.30 & 0.02 & 21.33 & 0.02 & 2.6--5.2 \\
\hline
\end{tabular}
\end{table}

J042911.69+270220.2 displayed a visual companion of similar luminosity and color located 1\arcsec\ north. However, the source is detected and reported in Bouy et al. (in prep.) ground-based images. It displays a proper motion measurement ($\mu_{\alpha} \cos{\delta}, \mu{\delta}$) = (6.9, 3.1) $\pm$ (3.6, 3.6) mas yr$^{-1}$ inconsistent with the primary’s motion of ($\mu_{\alpha} \cos{\delta}, \mu{\delta}$) = (6.9, -14.9) $\pm$ (2.5, 2.5) mas yr$^{-1}$, leading us to discard it as a background or foreground coincidence given that this difference in proper motion is too large to be orbital motion.

\begin{figure}
\centering
\includegraphics[width =0.48\textwidth]{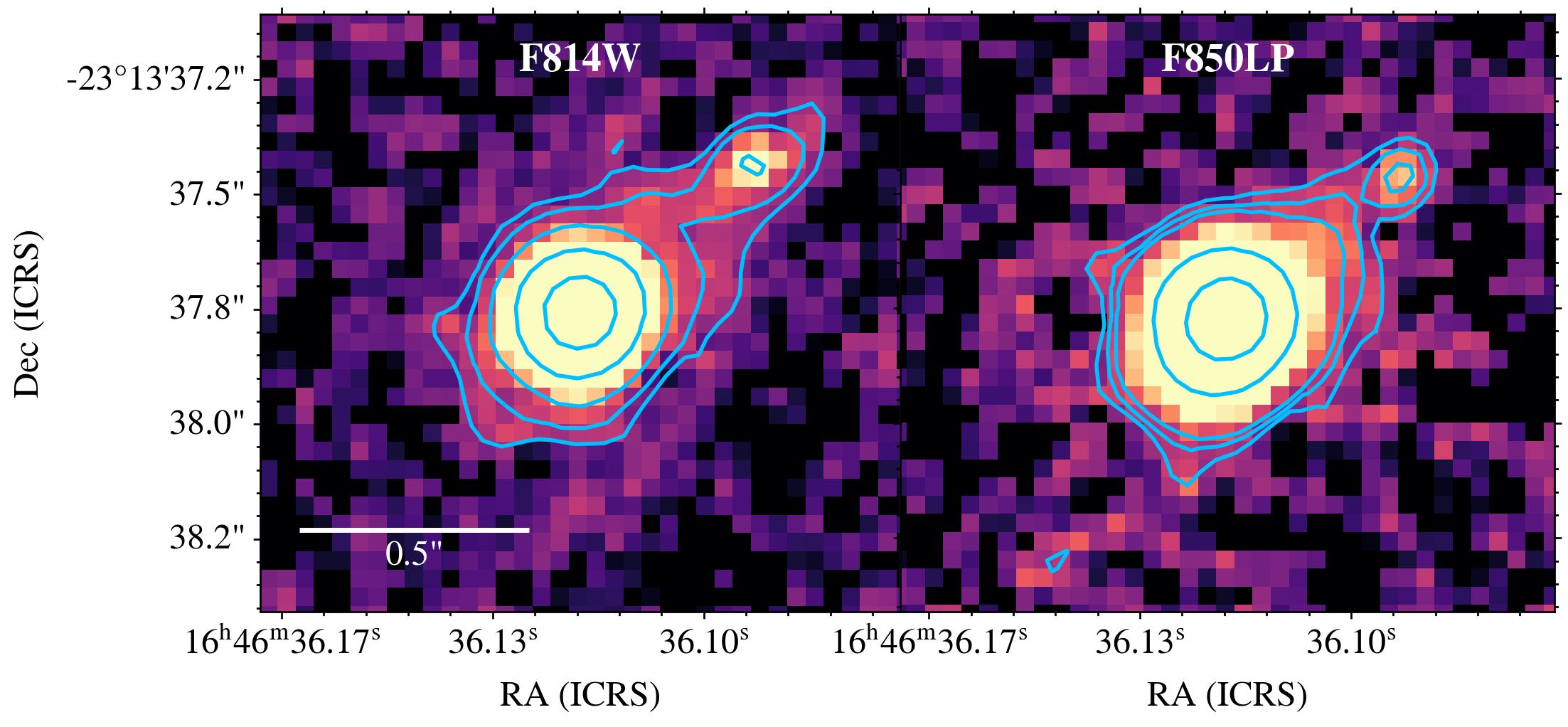}\\
\caption{HST F814W (left) and F850LP (right) images of the visual companion identified around J164636.12-231337.6 through direct inspection. A square root stretch is used. Contours highlight the extended emission originating from the visual companion. North is up and east is left.\label{fig:galaxy-comp}}
\end{figure}

J164636.12-231337.6 features a faint object (3\% of the target's brightness) at approximately 0\farcs5 separation, as shown in Fig.~\ref{fig:galaxy-comp}. The companion's color and luminosity, shown in Figure~\ref{fig:targets_cmd_uvis}, were significantly bluer than those of targets with similar luminosities. Additionally, the images revealed extended nebulosity roughly aligned with the companion’s position angle (Fig.~\ref{fig:galaxy-comp}). This, combined with its bluer color, suggested the companion could be a background galaxy, especially given the lower extinction in USco compared to Taurus. Based on these factors, the companion to J164636.12-231337.6 was rejected as a likely background galaxy.

We estimate the probability of finding a galaxy close to the 55 targets of our sample using the COSMOS 2020 catalog \citep{Weaver2022} that includes deep HST ACS observations in the F814W filter \citep{Koekemoer2007}. We used their \verb|lptype| star/galaxy discriminator to select only the galaxies, and computed the probability to find a galaxy within radii of 0\farcs5 and 1\farcs0 over the magnitude range between each target F814W magnitude and the limit of sensitivity of our images computed as described above and estimated at 26.5~mag in the F814W. We find that we expect to detect between $\sim$1 and $\sim$4 galaxies within radii of respectively 0\farcs5 and 1\farcs0 among our sample of 55 objects. 

To identify additional companions closer to the PSF core, we then applied three different PSF subtraction methods that take advantage of the remarkable stability provided by HST. 

\subsubsection{Least-squares fit with a natural PSF (PSFsub)}

The initial approach involved a straightforward least-squares adjustment. Since BDs and FFPs are significantly redder than most field stars and to avoid any chromatic effect, we opted to use images of our own targets as reference PSFs. Specifically, we selected the four brightest targets (see Table~\ref{tab:targets}) and used each of them as a PSF to subtract from all other targets. This strategy was designed to minimize systematic artifacts that could arise from a faint companion associated with an individual reference PSF. One drawback of using observations from different visits is the "breathing" effect of the HST--small variations in focus position that lead to subtle changes in the wings of the PSF \citep{Hasan1994}. As a result, for each target, we generated eight PSF-subtracted images, four per filter, and only considered candidate companions that are consistently identified using multiple PSF stars.

\subsubsection{\textit{Stra}KLIP}
\label{sec:straklip}

A second avenue of analysis uses principle component analysis to model and subtract the stellar PSFs with the software packages \straklip{}\footnote{\url{https://github.com/strampelligiovanni/StraKLIP}} \citep{Strampelli2022} and \publicwifi{} (described \ref{sec:public-wifi}). \straklip{} is a pipeline originally developed to detect and characterize close-in candidate companions to stars in HST/WFC3-IR and ACS datasets. For this study, its capabilities have been expanded to include the WFC3-UVIS dataset and to enable a more robust analysis of candidate companions.

Briefly, \straklip{} performs the following five fundamental steps:  

\begin{enumerate}
    \item {\bf Postage stamp creation:} Small \textit{postage-stamp} images (also referred to as \textit{tiles}) are generated for each source in the \textit{input catalog}. These images define the search regions, $\mathcal{S}$, that will be inspected for each target. $\mathcal{S}$ must be large enough to encompass the bright wings of the PSF while remaining small enough to avoid contamination from the wings of nearby catalog sources.  

    \item {\bf KLIP analysis:} PSF subtraction is performed for each source in each \textit{tile} using a set of local stars to construct a reference differential imaging (RDI) PSF library for the KLIP algorithm \citep{Soummer2012}.  

    \item {\bf Residual detections:} S/N maps of the PSF-subtracted \textit{residual tiles} are analyzed to identify previously undetected astronomical signals. To be classified as a candidate, a detection must meet two criteria: (i) it must appear at a minimum significance level of 5 S/N, and (ii) it must be present across multiple consecutive KLIP modes to rule out artifacts.  

    \item {\bf Candidate validation:} for non visual binaries, jack-knife tests are performed to insure that these candidates are not artifacts introduced by one of the reference images. This test is performed by subtracting the target with a PSF library where one at the time one reference is removed, and by analyzing each single residual. If the candidate disappears in one of them, that's a clear sign that was injected, and therefore not real.

    \item {\bf Candidate extraction:} PSFs forward modeling through the subtraction process \citep{Pueyo2016} are combined with Markov Chain Monte Carlo (MCMC) algorithms to extract validated candidate detections and determine key parameters, such as contrast, position angle, and separation relative to the primary target. Additionally, the sensitivity of the mass-separation parameter space is quantified through contrast curves, providing insights into occurrence rates. 
\end{enumerate}

\subsubsection{\publicwifi{}}
\label{sec:public-wifi}
\publicwifi{} (\textit{P}reviously \textit{U}ndiscovered \textit{B}inaries \textit{L}ying \textit{I}n \textit{C}lusters - \textit{Wi}de \textit{Fi}eld)\footnote{https://github.com/aggle/public\_wifi}, like \straklip{}, is a principal component analysis (PCA)-based faint companion detection pipeline designed to organize the analysis of large wide-field surveys of unocculted point sources. The main points of difference are in architecture -- \publicwifi{} is primarily object-oriented where \straklip{} is primarily database-oriented -- and in some specifics of implementation, e.g. how PSF forward modeling is implemented \citep{pyklip, Pueyo2016, Ruffio2017}. Though the two algorithms necessarily share many of the same basic steps described in \ref{sec:straklip}, they nevertheless provide semi-independent checks on each other with regard to companion detection and search depth.

\publicwifi{} is initialized with an input catalog containing, at minimum, columns for the target names, exposure identifiers, and the x- and y-coordinates of the target. If a target is detected in more than one exposure, each detection is recorded on a separate row, and further columns should be included to specify the differences between the exposures (e.g., a filter change or overlapping mosaic tile). The combination (target name, exposure identifier) should be unique, and is used to organize the PSF subtraction.
The \publicwifi{} pipeline creates an object for each target (internally called a \textit{Star}) that aggregates the relevant catalog rows, stores the metadata for all the associated exposures, and creates a postage stamp for each instance of the target in the survey. Once all \textit{Star} objects have been created, they are able to interface with each other to assemble reference PSF libraries for PCA-based PSF subtraction. \publicwifi{} currently uses the pyKLIP implementation of the KLIP algorithm \citep{Soummer2012}, but is well-suited to integrate other algorithms that combine images from a reference library to construct PSF models, such as LOCI \citep{Lafreniere2007} or NMF \citep{Ren2018}. Finally, wide-field surveys can contain tens or even hundreds of targets, presenting a significant challenge when it comes to synthesizing the analysis results in a sensible way. \publicwifi{} provides a useful browser-based data exploration interface for organizing, displaying and analyzing the resultant data products that allows the user to isolate each target and scroll through the related data products including the reference library used, post-subtraction residuals, SNR maps, matched filtering results, and more. A more complete description of \publicwifi{} will be provided in a future publication.

\subsubsection{Tentative candidates}
The three methods were employed to identify companion candidates and characterize their positions and fluxes. We note that \straklip{} and \publicwifi{} are based on similar fundamental principles and are expected to yield consistent results. Despite their similarities, we report the results from both methods, as they serve as an important sanity check and contribute to the robustness of our analysis.

\begin{figure}
\centering
\includegraphics[width =0.49\textwidth]{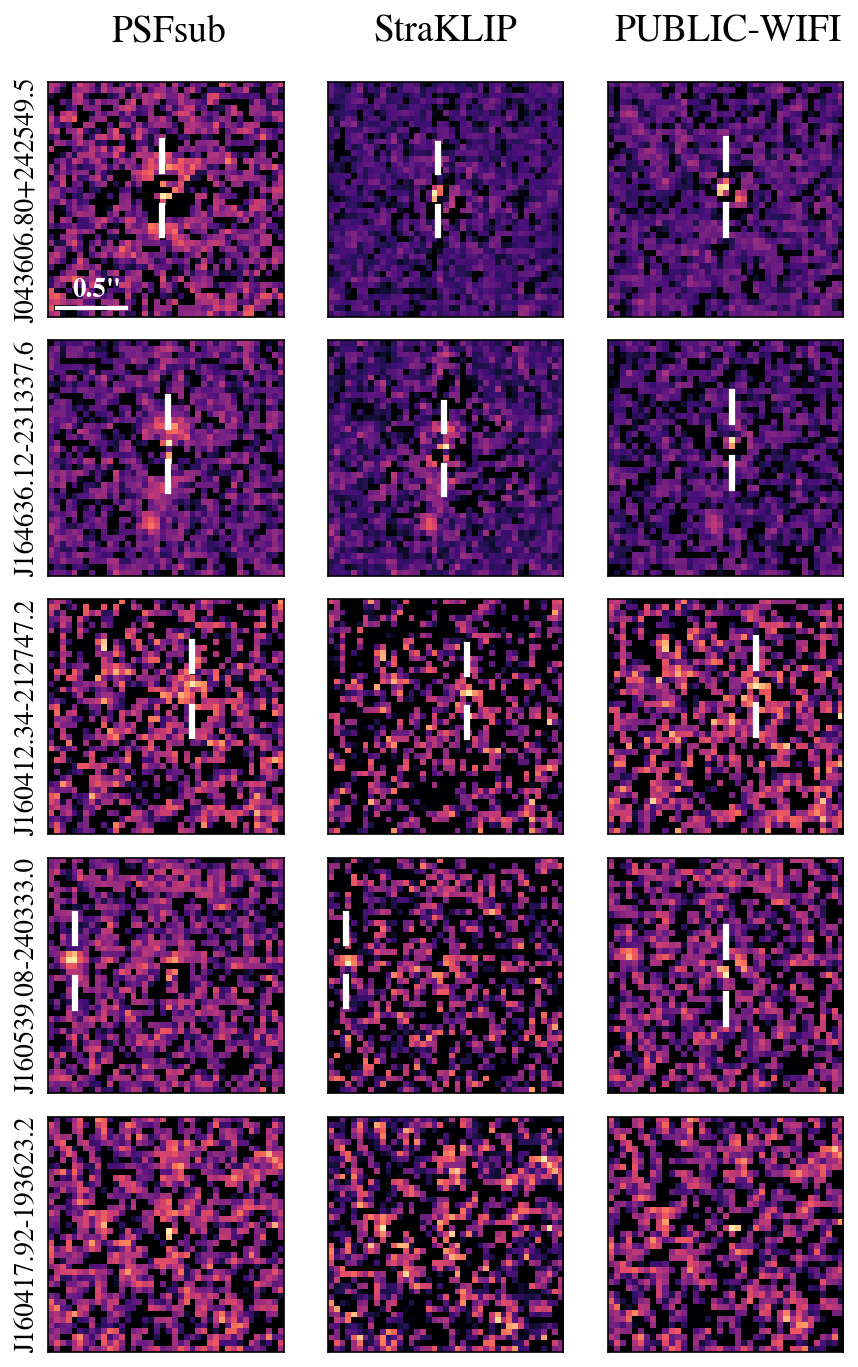}
\caption{Residuals for J043606.80+242549.5, J164636.12-231337.6, J160412.34-212747.2, and J160539.08-240333.0 in the F850LP filter, obtained using the three methods described in the text. The scale is indicated in the top-left image, with crosshairs marking the detection locations. 
For comparison, the residuals of an unresolved target (J160417.92-193623.2) are shown in the bottom row. The suspected galaxy is also visible in J164636.12-231337.6 images (see also Fig.~\ref{fig:galaxy-comp}). The orientation is in native detector coordinates.
\label{fig:residuals}}
\end{figure}

The three methods successfully recover J042705.86+261520.3 and identify 3 tentative detections in the F850LP filter around J043606.80+242549.5, J164636.12-231337.6, J160412.34-212747.2, and one clear detection at larger separation around J160539.08-240333.0. Their properties reported by \straklip{} are given in Table~\ref{tab:tentative}. Figure~\ref{fig:residuals} shows the residuals after modeling by the three methods. 

Two of these four new candidates exhibit relatively large separations, while all four have very small flux  ratios of just 3--5\%. If confirmed, these systems would be unusual compared to the known mass ratio distribution of slightly more massive ultracool dwarfs, which peaks around near-equal masses.

However, in some cases, the detections fall close to or even below the nominal 5-$\sigma$ threshold in one or several methods, and in two instances, they are at or below the Nyquist sampling limit, making them tentative. Due to their uncertain nature and the lack of F814W measurement, we do not include them in the subsequent analysis.

\begin{table}
\caption{Properties reported by \straklip{} for the faint companion candidates\label{tab:tentative}}
%\scriptsize
\centering
\begin{tabular}{rccc}
\hline
\hline
Object & $\delta$ & P.A. & Flux ratio \\
       & (mas)    & (\degr) & (F850LP) \\
\hline
J160539.08-240333.0 & 701.3 & 225 & 0.03 \\
J164636.12-231337.6 & 65.6  & 151 & 0.04 \\
J160412.34-212747.2 & 227.6 & 106 & 0.05 \\
J043606.80+242549.5 & 85.6  & 166 & 0.05 \\
\hline
\end{tabular}
\end{table}

\subsection{VLT images}
Detecting companions in AO images is significantly more challenging than in HST images. The shape of the point-spread function (PSF) is influenced not only by the seeing conditions but also by the brightness and distance of the wavefront reference star from the target. In our case, all targets were indeed too faint for direct wavefront sensing, making the observations particularly susceptible to tip-tilt anisoplanatism. As a result, most objects appear slightly elongated in the direction of the tip-tilt reference star. Furthermore, the observations were conducted in queue mode, spread over multiple nights, leading to substantial variations in image quality between nights and even within individual nights. Given that the AO correction was generally moderate, we did not attempt PSF subtraction to search for faint companions within the PSF core as was done for the HST data, because constructing a reliable PSF model for subtraction was not feasible.

\begin{figure}
\centering
\includegraphics[width =0.49\textwidth]{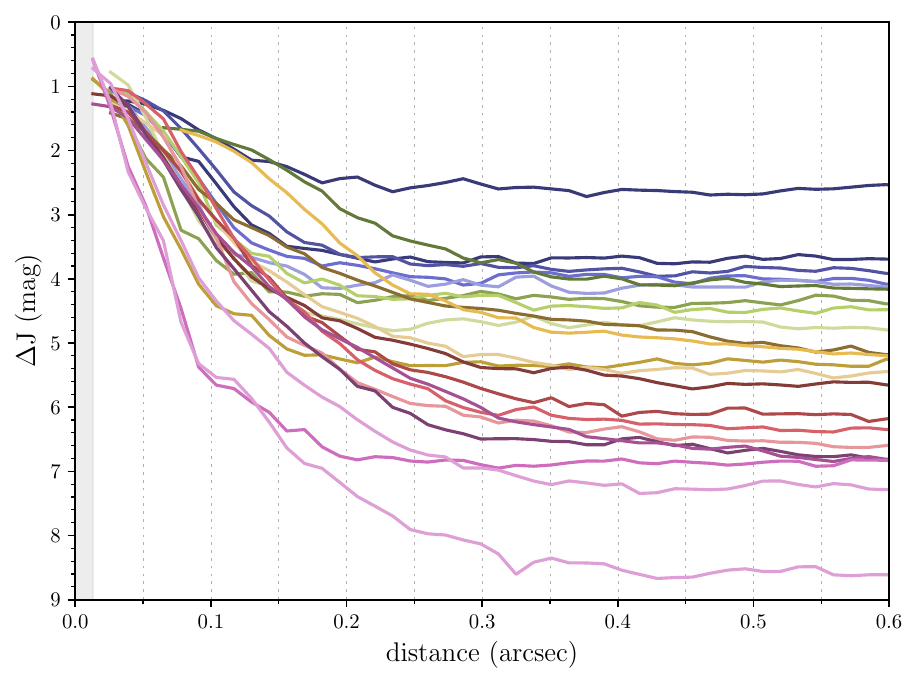}
\caption{Same as Fig.~\ref{fig:contrast_hst} for the VLT ERIS J-band images, computed using 2-pixel (26~mas) increments. }
\label{fig:contrast_eris}
\end{figure}

Instead, we limited our analysis to a visual inspection of the images. Figure~\ref{fig:contrast_eris} shows that, in most cases, our observations should have also been sensitive enough to detect companions with a flux ratio of $\sim$0.5 ($\Delta$J=0.75~mag) at separations as small as 0\farcs05 (or approximately 7~au at the distance of Taurus and USco). This sensitivity is comparable to that achieved in previous studies of the same association for similar ultracool objects by \citet{Bouy2006, Kraus2006, Biller2011, Kraus2012} and to the HST images presented above. This inspection did not reveal any companions within 0\farcs6 of the 22 targets. Companions at larger separations would have been detected and reported in \citet{MiretRoig2022} as common proper-motion companions. 

\begin{figure}
\centering
\includegraphics[width =0.49\textwidth]{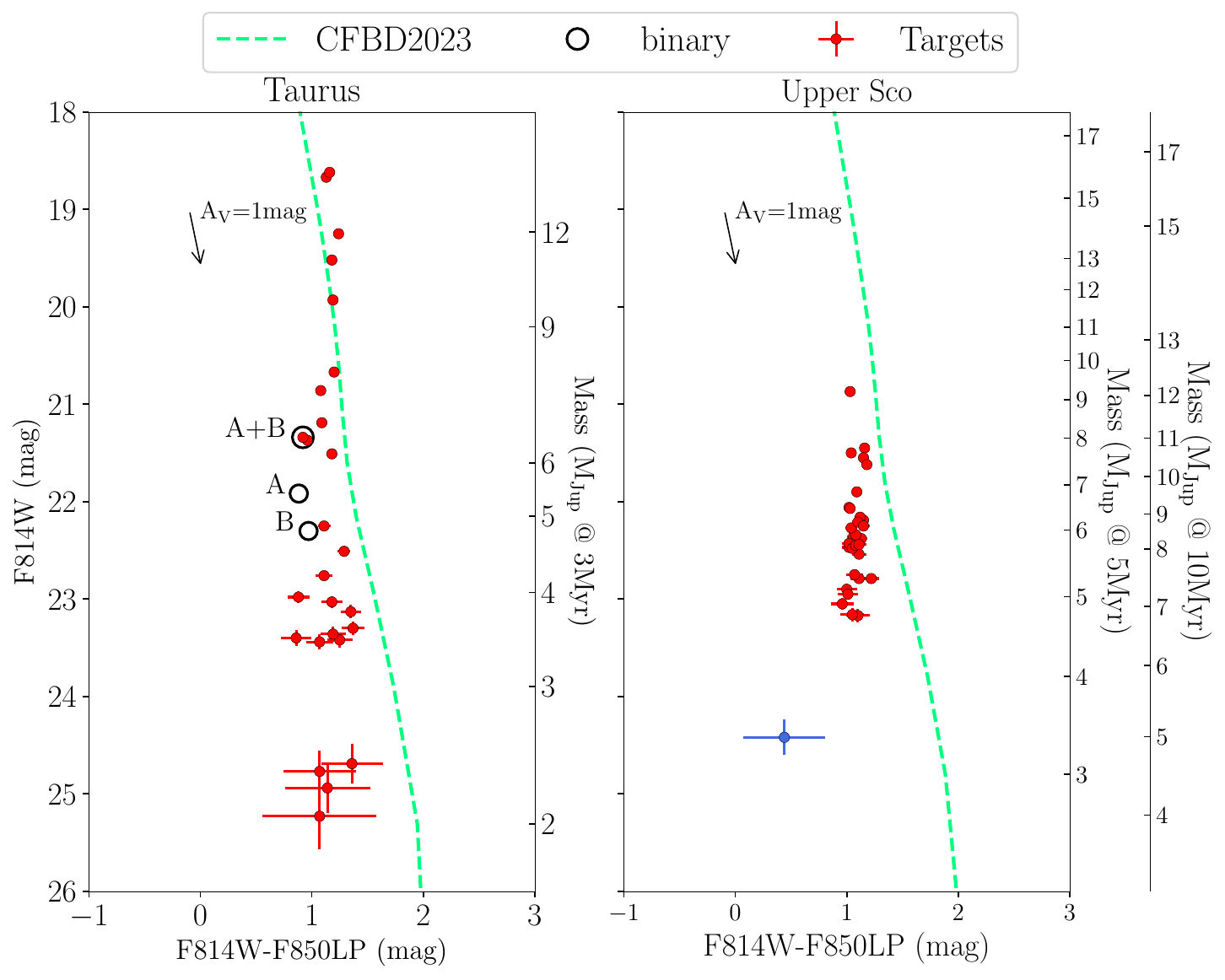}
\caption{(F814W, F814W-F850LP) diagram of Taurus targets (left panel) and USco targets (right). The \citet{Chabrier2023} isochrones at resp. 3~Myr (Taurus) and 5 and 10~Myr (USco) and 140pc are represented (green line) and the corresponding masses are indicated  on the right vertical axis. The 5 and 10~Myr isochrones overlap almost perfectly and are represented as one. The combined photometry (A+B) and individual components (A, B) of the binary candidate J042705.86+261520.3 are overplotted as labeled black open circles. The suspected background galaxy located 0\farcs5 from J164636.12-231337.6 is indicated as a blue dot. A reddening vector A$_{\rm V}$=1~mag is also represented. 
\label{fig:targets_cmd_uvis}}
\end{figure}

%--------------------------------------------------------------------
\section{Discussion}

\subsection{Multiplicity statistics}
In this section we discuss the result obtained with the HST survey only, which is homogeneous in sensitivity and spatial resolution. The detection of 1 candidate among a sample of 55 objects leads to an overall binary fraction of 1.8$^{+2.6}_{-1.3}$\%, computed using Bayesian inference for a binomial proportion. This number is consistent with the values reported for slightly more massive objects in the field \citep{Burgasser2006, Fontanive2018, Fontanive2023}. 

With only one detected binary, it is not possible to derive meaningful statistics on the distributions of separation and mass ratio. Nevertheless, we note that we did not identify any wide\footnote{\emph{wide} meaning with a separation larger than $\gtrsim$30~au corresponding to 5-$\sigma$ of the separation distribution reported for field late-T and Y dwarfs \citep{Fontanive2018}} binaries among the HST targets and VLT targets below the planetary mass limit, compatible at 1-$\sigma$ credible interval with a wide binary fraction of up to $\le$1.8\%\footnote{\label{note1} determined using Bayesian inference for a binomial proportion}. The separation of the binary candidate is close to the resolution limit of HST observations, suggesting that we may be missing the bulk of the FFP binary population. Although challenging with current instrumentation, higher spatial resolution observations are needed to explore the currently unprobed separation range below 7~au. We also note that the candidate exhibits a relatively "common" and near-equal mass ratio compared to the distribution observed for older and slightly more massive objects in the field \citep{Burgasser2006, Fontanive2018, Fontanive2023}.

\begin{figure*}
\centering
\includegraphics[width =\textwidth]{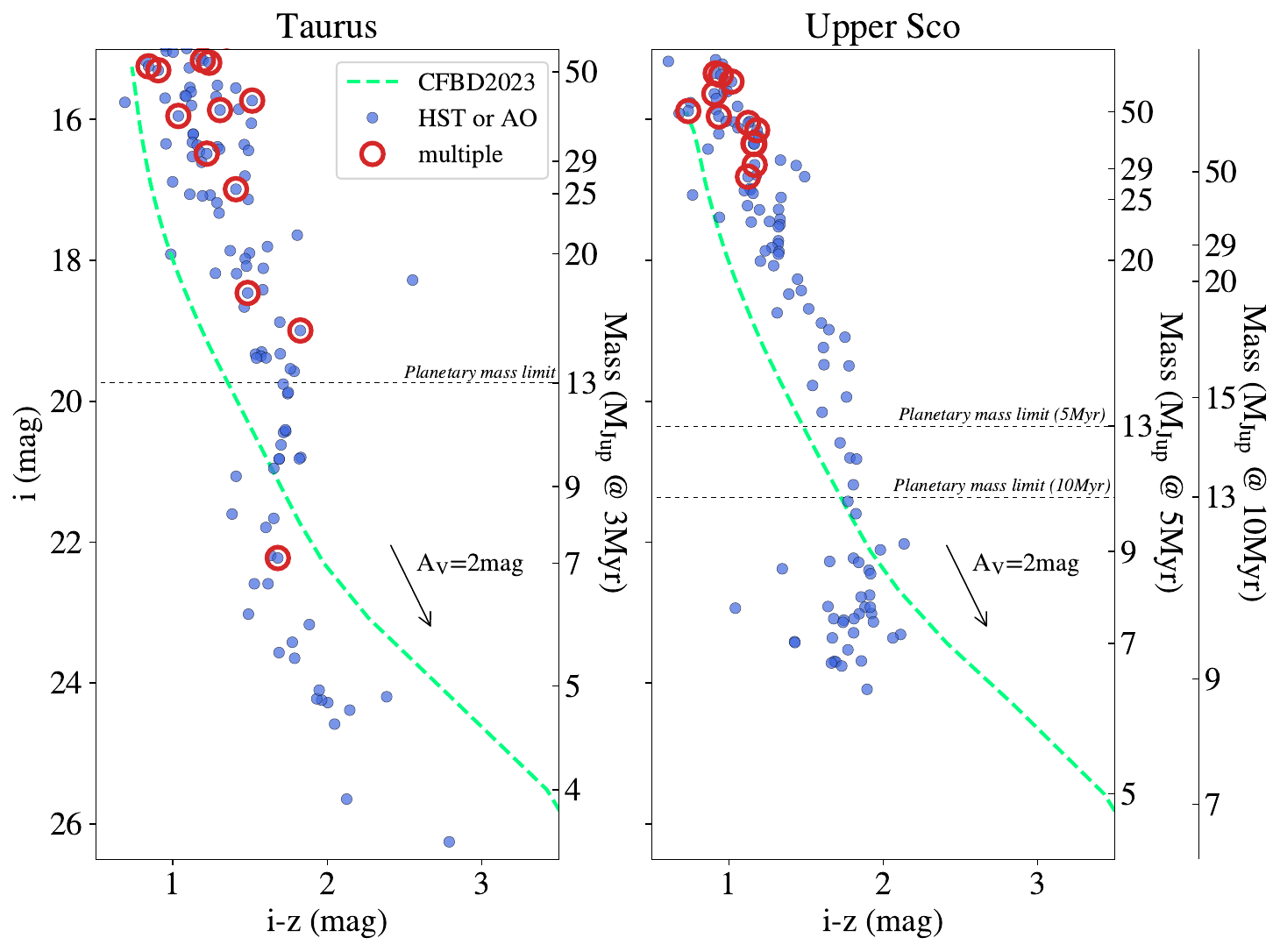}
\caption{ ({\it i, i-z}) diagram of Taurus members (left panel) and USco members (right) observed at high spatial resolution using either HST or adaptive optics (blue dots) by \citet{Todorov2010,Todorov2014,Bouy2006,Biller2011,Kraus2012,Kraus2006,Konopacky2007} as well as this work. The \citet{Chabrier2023} isochrones at 3~Myr (Taurus) and 5 and 10~Myr (USco) 140pc are represented by a green dashed line and the corresponding masses are indicated on the right vertical axis. The 5 and 10~Myr isochrones overlap almost perfectly and are represented as one in the USco panel. Resolved binaries are over-plotted as red open circles. A reddening vector A$_{\rm V}$=2~mag is also represented. See also Table~\ref{tab:literature-usco} and \ref{tab:literature-taurus}. 
\label{fig:comparison_taurus_usco}}
\end{figure*}

\subsection{Dependence on environmental conditions}
If considering the Taurus sample only, from which the binary candidate comes from, the binary fraction rises to 3.8$^{+5.2}_{-2.8}$\% in Taurus while the lack of companion candidate  among 30 objects in USco is  compatible at 1-$\sigma$ credible interval with a binary fraction of up to $\le$3.6\%\footnotemark[3]. The difference between the two regions is not statistically conclusive mostly because of the limited sample sizes, but it suggests a potential trend that warrants further investigation with larger samples. 

In this context, we complemented the present work by compiling a comprehensive list of all low-mass sources in Taurus and Upper Sco that have been observed at high spatial resolution with either HST or AO over the past two decades. Both regions have been extensively surveyed, yielding statistically significant samples ($\sim$100 sources each) that reach down to a few Jupiter masses in each association. Tables~\ref{tab:literature-taurus} and~\ref{tab:literature-usco} list the Taurus and Upper Sco members reported in the literature in \citet{Todorov2010, Todorov2014, Bouy2006, Biller2011, Kraus2012, Kraus2006, Konopacky2007, Luhman2007} and \citet{Luhman2009}. The corresponding $(i,\, i-z)$ colour--magnitude diagrams are shown in Fig.~\ref{fig:comparison_taurus_usco}.

Some sources appear in multiple studies, occasionally under different designations. These duplications were identified and accounted for when compiling the final samples presented in Tables~\ref{tab:literature-taurus} and~\ref{tab:literature-usco}.

The spatial resolution and sensitivity of these HST and AO studies are variable but comparable, typically achieving  flux ratios of $\sim$0.5 or better at separations of about 0\farcs05$\sim$0\farcs08. Since Taurus and Upper Sco lie at comparable distances ($\sim$140~pc), the high--spatial-resolution surveys performed in both regions reach very similar sensitivities to BDs, FFPs, and their companions. The interpretation of non-detections in USco, however, depends on the assumed age of that association. Given the uncertainty and spread on the ages of the USco sub-groups mentioned above, we consider two commonly adopted age scenarios: 5~Myr and 10~Myr bracketing most of the sub-groups present in that region. 

(i) Case 1: If USco is assumed to be 5~Myr old, the faintest magnitude at which no companions have been detected ($i \ge 16.8$~mag) corresponds to a mass limit of 29~M$_{\rm Jup}$ according to the evolutionary models of \citet{Chabrier2023}. At the younger age of 3~Myr appropriate for Taurus, this same mass corresponds to a limit of $i \ge 16.6$~mag. Among the Taurus objects surveyed at high spatial resolution, 4 binaries are found among 67 sources with $i \ge 16.6$~mag (see Table~\ref{tab:literature-taurus}), yielding a binary fraction of $5.6^{+3.2}_{-2.3}$\%. With the exception of our new candidate (which still lacks spectroscopic confirmation), these Taurus objects have reported spectral types between M5.5 and M8.5 and therefore cannot be explained as more massive embedded objects.

(ii) Case 2: If Upper Sco is instead assumed to be 10~Myr old, the non-detection limit of $i \ge 16.8$~mag translates to a higher mass threshold of 50~M$_{\rm Jup}$. At 3~Myr in Taurus, this mass corresponds to $i \ge 15.3$~mag. In this regime, 8 binaries are identified among 94 Taurus objects fainter than this limit, corresponding to a binary fraction of $7.8^{+3.0}_{-2.4}$\%. Again, their reported spectral types (later than M5.25) are consistent with masses below approximately 50~M$_{\rm Jup}$, and thus these systems are not embedded more massive interlopers.

We note that even if we conservatively exclude our newly reported candidate until spectroscopic confirmation is available, the corresponding fractions change only marginally to $4.3^{+2.9}_{-2.0}$\% (5~Myr/29~M$_{\rm Jup}$ regime) and $7.0^{+2.8}_{-2.3}$\% (10~Myr/50~M$_{\rm Jup}$ regime). Similarly, \citet{Kerr2023} found that the multiple system [MDM2001] CFHT-BD-Tau 18 (i=15.73~mag) might be a member of the 11~Myr SCYA~108 group. If we conservatively exclude it as well, the corresponding fractions change marginally only when assuming 10~Myr to $6.1^{+2.7}_{-2.1}$\% (50~M$_{\rm Jup}$ regime).

In contrast with the several detections in Taurus, no visual binary candidates have been found among the 80 USco members with $i \ge 16.8$~mag observed at high spatial resolution. This implies an upper limit on the binary fraction of $\le 1.2$\% in USco,\footnotemark[3] significantly lower than in Taurus for either of the two ages (respectively mass regimes) considered. These results therefore suggest a genuine and statistically significant difference in the multiplicity properties of objects below approximately 30--50~M$_{\rm Jup}$ between the two regions.

We also note that the Taurus fraction mentioned above should be considered as a lower limit: while the multiple systems have observed spectral types consistent with masses below approximately 30--50~M$_{\rm Jup}$, the other sources could include a few more massive embedded interlopers that might decrease the final binary fraction in the correspond mass range.

At brighter magnitudes ($i < 16.8$~mag), the relative frequencies of multiple systems in Taurus and Upper Sco seem statistically indistinguishable, consistent with previous high-resolution studies \citep{Bouy2006, Biller2011, Kraus2012, Kraus2006, Konopacky2007}.

\subsection{Implications for the formation process of BDs and FFPs}

The Taurus molecular clouds and Upper Sco were selected for this study because they represent drastically different star-forming environments \citep{2008hsf1.book.....R,2008hsf2.book.....R}. Upper Sco is a rich OB association hosting several massive B stars and approximately 3\,500 members \citep{Blaauw1946,MiretRoig2022}, whereas Taurus is a sparse  association devoid of massive stars, with about 900 low-mass members distributed along filamentary molecular clouds \citep{Galli2019,Joncour2018}. Upper Sco is also moderately older, with subgroups spanning ages from $\sim$3 to 20~Myr and a median age around 10~Myr \citep{MiretRoig2022b,Luhman2022,Schmitt2022,Zerjal2023,Ratzenbock2023}, compared to typical ages of 1--3~Myr for the Taurus molecular clouds \citep{Luhman2023}.

If real, the apparent lack of binary systems below $\sim$30--50~M$_{\rm Jup}$ in Upper Sco may therefore reflect differences in formation pathways between the two regions. A variety of mechanisms have been proposed for the formation of low-mass brown dwarfs and free-floating planets, including dynamical processes such as planet--planet or brown dwarf--brown dwarf scattering \citep{Veras2012,lazzoni2024}, stellar flybys \citep{Wang2024}, photo-erosion of fragmenting stellar cores \citep{diamond2024}, or grazing encounters between protoplanetary disks \citep{Fu2025}. More dynamically active formation channels are generally expected to produce fewer bound systems than quiescent collapse scenarios, and it is plausible that the dominant mechanisms operating in Upper Sco and Taurus differ. Alternatively, the efficiency of forming very low-mass objects through core collapse may itself be reduced in Upper Sco compared to Taurus. In particular, producing extremely low-mass prestellar cores may be more difficult in environments subject to stronger radiative and mechanical feedback from massive stars and their supernovae \citep{Preibisch1999}.

A physically motivated explanation for these differences may be linked to the thermal properties of the parental molecular clouds and their impact on fragmentation. Both the characteristic mass scale of fragments and their typical separations are expected to depend sensitively on the cloud temperature through the Jeans formalism. Since the Jeans mass scales approximately as $M_{\rm Jeans} \propto T^{3/2}$ \citep[e.g. Eq.~14 in][]{Palau2024}, implying that warmer molecular clouds fragment at systematically higher minimum masses (assuming similar densities for opacity-limited fragmentation in all clouds). As a result, the production of very low-mass objects through core collapse, including low-mass BDs and FFPs, is expected to be suppressed in regions with elevated cloud temperatures. \citet{Palau2024} argue that this effect explains the paucity of proto--brown dwarfs near the planetary-mass limit in Ophiuchus (compared to Taurus, see their Fig.~7 and Sect.~7.4), a region known to host relatively warm and dense gas \citep{Alves2025}. Given the presence of massive stars in Upper Sco, it is likely that its parental clouds were significantly warmer than those in Taurus, which could naturally account for the reduced number of companions in the planetary-mass regime.

In addition to influencing the characteristic mass scale, the cloud temperature also affects the typical spatial scale of fragmentation. The Jeans length, which represents the minimum size of a perturbation capable of gravitational collapse, scales as $L_{\rm Jeans} \propto T^{1/2}$ \citep[][]{Bontemps2010}. Higher temperatures therefore lead to larger initial separations between fragments. For instance, at a density of $n \sim 10^9$~cm$^{-3}$, a cloud at $T \sim 40$~K \citep[typical of dense gas as found in Ophiuchus and Orion, ][]{Alves2025, Tang2018} yields a Jeans length of approximately 280~au, compared to about 140~au at $T \sim 10$~K \citep[typical of the Taurus molecular clouds, ][]{delBurgo2005}. Such a factor of two to three increase in characteristic separation may be sufficient to prevent newly formed low-mass fragments from remaining gravitationally bound, thereby reducing the likelihood of forming binary systems.

Numerical simulations further illustrate the strong sensitivity of multiplicity outcomes to the initial cloud temperature. \citet{Riaz2014} showed that, for modest azimuthal density perturbations, increasing the initial temperature above $\sim$10~K can suppress binary formation altogether, leading instead to a single low-mass fragment. Conversely, small temperature variations below 10~K can result in changes of order 100~au in the separation of binary components. These results demonstrate that relatively small differences in the thermal conditions of star-forming clouds can have a decisive impact on the formation and survival of low-mass binary systems.

Taken together, these considerations suggest that the absence of binaries below $\sim$50~M$_{\rm Jup}$ in Upper Sco may reflect intrinsically different initial conditions compared to Taurus, rather than observational biases alone. Warmer parental clouds would both inhibit the formation of low-mass BD and FFPs via core collapse, the main mechanism to produce binaries, and favor wider initial separations, reducing the probability that such low-mass fragments remain bound. While this interpretation remains speculative, it provides a physically motivated framework that can be tested with future observational constraints on cloud temperatures and with spectroscopic confirmation of candidate systems.

This interpretation is further supported by the discovery of wide binary BD and FFP candidates in several young, nearby, and relatively massive star-forming regions, where warmer initial conditions are expected. In the Orion Nebula Cluster (ONC), a dense and massive environment with an age of $\sim$1~Myr, \citet{Luhman2024b} reported one candidate binary FFP ([LAB2024]~139) among a sample of 242 brown dwarfs and planetary-mass objects. This system has a wide projected separation of 132~au and a small mass ratio of $q \sim 0.13$. Similarly, \citet{Langeveld2024} identified two binary candidates among six FFP candidates in the 1--3~Myr NGC~1333 cluster. Both systems (NGC1333-NN10 and NGC1333-NN12) exhibit wide separations of 164~au and 76~au, respectively, with relatively high mass ratios of $q \sim 0.7$--0.8.

\begin{figure}[!h] %{R}{0.65\textwidth}
\centering
\includegraphics[width =0.49\textwidth]{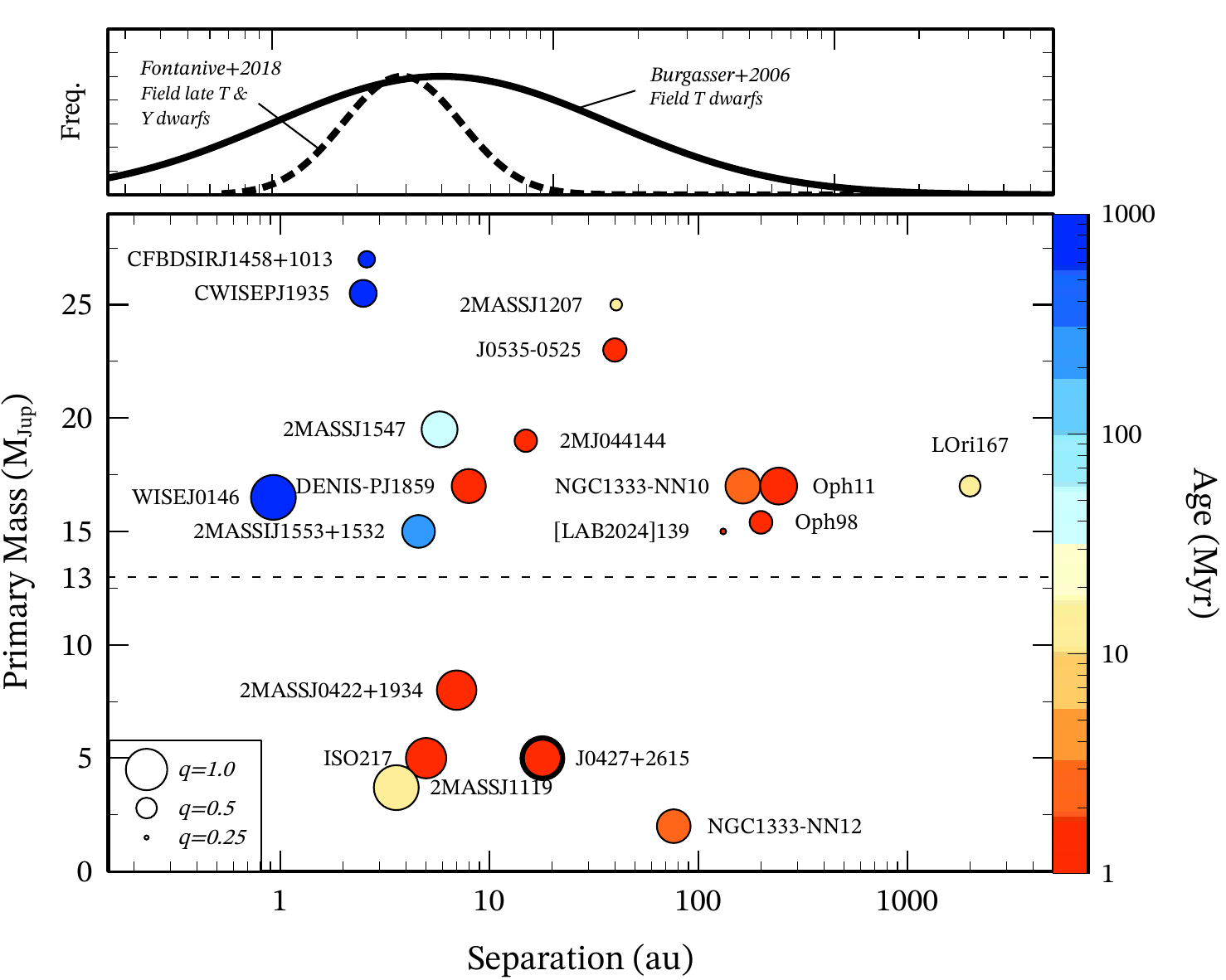}
\caption{Lower panel: Primary mass vs separation for known BD-FFP and FFP-FFP binaries or candidates. The color scale indicates the age and the size is proportional to the mass ratio. 2MASSJ~1119 data from \citet{Best2017}. WISEJ0146 data from \citet{Dupuy2015}. 2MASSJ044144, ISO217 and 2MASSJ0422 data from \citet{Todorov2014}. DENIS-PJ1859 data from \citet{Bouy2004}. Data for 2MASSJ~1547 from \citet{Calissendorff2019}. Data for 2MASSJ J1553 from \citet{Dupuy2012}. Data for NGC1333-NN10 from \citet{Langeveld2024}. Data for [LAB2024] 139 from \citet{Luhman2024b}. Data for Oph~98 from  \citet{Fontanive2020} and for Oph~11 from \citet{Close2007}. Data for LOri~167 from \citet{Barrado2007}. 2MASS1207 data from \citet{Chauvin2004}. CFBDSIRJ1458+1013 data from \citet{Delorme2010}. CWISEPJ1935 data from \citet{DeFurio2025b} J0535-0525 data from \citet{deFurio2022}. The new multiple system from the present study (J042705.86+261520.3) is represented with a circle with a thick edge and follows the same color and size scales. The upper panel shows the distribution of separations reported for older and slightly more massive field T-dwarfs from  \citet{Burgasser2006} (solid line) and field late-T and Y dwarfs from \citet{Fontanive2018} (dashed line).
\label{fig:binaries}}
\end{figure}

If confirmed as physical pairs, these systems would join the small population of young, wide binary systems with components at or near the planetary-mass regime, including Oph~11 \citep[$\rho = 243$~au;][]{Close2007}, Oph~98 \citep[$\rho = 200$~au;][]{Fontanive2020}, the extremely wide LOri~167 system in the Collinder~69 cluster \citep[$\rho \sim 2\,000$~au;][]{Barrado2007}, and the benchmark 2MASS~J1207-3932 system in the TW~Hya association \citep{Chauvin2004}, illustrated in Fig.~\ref{fig:binaries}. Similar weakly bound and slightly more massive wide BD binary candidates have also been reported in Upper Sco \citep{Bouy2006,Bejar2008} and in Orion \citep{deFurio2022}. In contrast, only one wide BD binary has been reported to date in Taurus (FU~Tau~B; \citealt{Luhman2009}), which is coincidentally located in the vicinity of $\chi$-Tau (B9V) and  62~Tau (B3V), two of the very few B-stars present in Taurus. However, most of these wide systems have not yet been confirmed as physical pairs and, given the stellar densities of these clusters, they could instead be chance projections of two unbound members or even unrelated background sources. Confirmation of their nature is therefore necessary before drawing any firm conclusion. But if confirmed, the existence of wide systems in massive and dynamically active regions and lack thereof in Taurus is qualitatively consistent with expectations from fragmentation in warmer parental clouds, where larger Jeans lengths naturally lead to wider initial separations. In this context, the lack of detected binaries below $\sim$30--50~M$_{\rm Jup}$ in Upper Sco may not reflect an absence of multiplicity altogether, but rather the absence of close and moderately bound systems in this mass regime. Low-mass binaries formed with intrinsically wide separations would be more difficult to detect and more susceptible to disruption, naturally reducing the observed binary fraction at small separations in warmer clouds.

Finally, dynamical evolution may further accentuate these differences. Even if wide, low-mass binaries form efficiently in dense or massive associations, their low binding energies make them particularly vulnerable to early disruption through interactions with other members or with the evolving cluster potential. As a result, the observed multiplicity properties in Upper Sco likely reflect a combination of initial formation conditions and subsequent dynamical processing, whereas the colder and sparser environment of Taurus may favor the formation and long-term survival of closer, bound systems in the planetary-mass regime.

%--------------------------------------------------------------------
\section{Conclusion}
We present high-spatial-resolution observations using HST and VLT of a sample of 38 FFPs and 14 low-mass BD members of the USco association, as well as 25 FFP members of the Taurus association. We detect one binary candidate in Taurus with a separation of 111.9$\pm$0.4~mas corresponding to 18~au at 160~pc, a mass ratio of 0.9 and estimated component masses in the range 3--6~\mjup for the primary and 2.6--5.2~\mjup for the secondary, according to \citet{Chabrier2023} models and assuming no extinction. Spectroscopic observations are required to confirm the nature of this pair. No binary candidates were identified in USco and no wide binary candidates were identified in either association. Using advanced PSF-fitting methods, we also report the tentative detection of four very faint companion candidates in the F850LP filter, with flux ratios of only 3\% to 5\%. Given the uncertain nature of these detections, which are at the resolution and/or detection limits of the HST images, these candidates should be considered with caution and require confirmation.

When combined with previous high-spatial-resolution surveys of the very low-mass populations in these regions, our findings reveal a statistically significant difference in the multiplicity properties below 30--50~\mjup, with a binary fraction of $4.5^{+3.1}_{-2.1}$\% below 29\mjup and $8.9^{+3.3}_{-2.7}$\%  below 50\mjup in Taurus and an upper limit of $\le$1\% in USco over the corresponding luminosity range.

The absence of visual binaries below $\sim$30–50~M$_{\rm Jup}$ in Upper Sco, contrasted with several detections in Taurus over the same mass range, may reflect intrinsically different formation conditions rather than observational biases alone. Upper Sco and Taurus probe markedly distinct star-forming environments: Taurus formed in cold, sparse molecular filaments, whereas Upper Sco originated in a more massive, warmer cloud complex influenced by early massive stars. In such warmer environments, the Jeans mass is expected to increase and the characteristic fragmentation scale to shift toward larger separations, naturally suppressing the formation of the lowest-mass bound systems and favoring wider, more weakly bound pairs. This interpretation is qualitatively consistent with the existence of a few wide brown dwarf and planetary-mass binaries reported in Upper Sco, Orion, and other massive associations, and with the relative scarcity of close, low-mass systems in these regions. While dynamical disruption may further reduce the survivability of the most weakly bound binaries, especially in dense clusters, the lack of close companions at the lowest masses in Upper Sco is likely rooted primarily in the initial fragmentation process. Although speculative at this stage, this environmentally driven picture provides a physically motivated framework to interpret the observed multiplicity differences and can be tested with future spectroscopic confirmation of candidates, improved constraints on cloud temperatures, and deeper high-resolution surveys across a broader range of star-forming environments.

These results highlight the sensitivity of low-mass BD and FFP formation to inborn and evolutionary effects. They raise new questions about the exact role of environmental factors in shaping the low-mass BD and FFP population and underlines the need for further observational confirmation of the physical nature of these wide systems. Additionally, future observations at higher spatial resolution, particularly targeting the unexplored separation range below 7~au, will be crucial to providing a more complete picture of how environmental factors contribute to the formation and evolution of low-mass objects.

\begin{acknowledgements}
We thank the referee for a prompt and thorough review that has significantly improved the quality and clarity of the manuscript. We are grateful to Tricia Royle for her support with the HST observations.  Based in part on observations with the NASA/ESA Hubble Space Telescope obtained at the Space Telescope Science Institute, which is operated by the Association of Universities for Research in Astronomy, Incorporated, under NASA contract NAS5-26555. Support for Program number HST-GO-17167 was provided through a grant from the STScI under NASA contract NAS5-26555. A.P. acknowledges financial support from the UNAM-PAPIIT IN120226 grant, M\'exico. JO acknowledge financial support from "Ayudas para contratos postdoctorales de investigación UNED 2021". DB and NH have been funded by grant No.  PID2023-150468NB-I00 by the Spain Ministry of Science, Innovation/State Agency of Research MCIN/AEI/ 10.13039/501100011033 and by “ERDF A way of making Europe”. PABG acknowledges financial support from the São Paulo Research Foundation (FAPESP) under grant 2020/12518-8. 
This research has made use of the VizieR catalogue access tool, CDS, Strasbourg, France. The original description of the VizieR service was published in \citet{vizier}. This research made use of Photutils, an Astropy package for detection and photometry of astronomical sources \citep{photutils}.
This work made use of GNU Parallel \citep{Tange2011a}, astropy \citep{astropy2013,astropy2018}, Topcat \citep{Topcat}, matplotlib \citep{matplotlib}, Plotly \citep{plotly}, Numpy \citep{numpy}, aplpy \citep{aplpy2012,aplpy2019}.
\end{acknowledgements}


\bibliographystyle{aa} % style aa.bst

\begin{table*}
\scriptsize
\caption{HST Targets \label{tab:targets}}
\begin{tabular}{cccccccccccccccc}
\hline\hline
  Object ID & RA (J2000) & Dec (J2000) & Mass & i & $\sigma_i$ & z & $\sigma_z$ & F814W & $\sigma_{F814W}$ & F850LP & $\sigma_{F850LP}$ & J & $\sigma_J$ \\
  & (deg) & (deg) & (M$_{\odot}$) & (mag) & (mag) & (mag) & (mag) & (mag) & (mag) & (mag) & (mag) & (mag) & (mag)  \\
\hline  
\multicolumn{14}{c}{Taurus} \\
\hline
J041514.73+280009.0 & 63.81145 & 28.00243 & 0.021 & 19.88 & 0.05 & 18.13 & 0.05 & 18.67 & 0.01 & 17.54 & 0.01 & 15.07 & 0.05 \\
J041546.22+280811.4 & 63.94261 & 28.13649 & 0.005 & 26.26 & 0.21 & 23.47 & 0.05 & 24.77 & 0.21 & 23.7 & 0.25 & 19.96 & 0.14 \\
\multicolumn{14}{c}{$\cdots$}\\
J041548.09+291132.9 & 63.95039 & 29.19245 & 0.021 & 20.62 & 0.05 & 18.92 & 0.05 & 19.52 & 0.01 & 18.34 & 0.01 & 15.7 & 0.05 \\
J041947.39+281534.6 & 64.94751 & 28.25954 & 0.007 & 23.02 & 0.05 & 21.53 & 0.05 & 22.25 & 0.04 & 21.14 & 0.04 & 18.15 & 0.05 \\
\hline  
\multicolumn{14}{c}{Upper Sco} \\
\hline
J154540.24-242207.2 & 236.41761 & -24.36874 & 0.008/0.010 & 23.41 & 0.1 & 21.99 & 0.07 & 22.9 & 0.06 & 21.9 & 0.07 & 18.55 & 0.06 \\
J154915.32-244139.1 & 237.31381 & -24.69425 & 0.008/0.010 & 22.92 & 0.05 & 21.27 & 0.05 & 22.07 & 0.03 & 21.04 & 0.04 & 18.2 & 0.06 \\
\multicolumn{14}{c}{$\cdots$}\\
J155150.21-213457.4 & 237.95917 & -21.58269 & 0.008/0.011 & 23.09 & 0.05 & 21.28 & 0.05 & 22.27 & 0.04 & 21.23 & 0.04 & 18.21 & 0.06 \\
J155416.68-263018.1 & 238.56942 & -26.50513 & 0.009/0.015 & 22.27 & 0.05 & 20.62 & 0.05 & 21.5 & 0.02 & 20.46 & 0.02 & 17.58 & 0.05 \\
\hline
\end{tabular}
\tablefoot{Full table available in electronic format.  Masses estimated assuming 3~Myr in Taurus and 5 and 10~Myr in Upper Sco.}
\end{table*}

\begin{table}
\caption{VLT targets \label{tab:vlt-targets}}
\tiny
\begin{tabular}{rcccccccccc}
\hline
\hline
  Object ID &RA (J2000) &Dec (J2000)  &i &$\sigma_i$ &z &$\sigma_z$ &J &$\sigma_J$ &       strehl & Mass \\
     & (deg) & (deg)  & (mag) & (mag) & (mag) & (mag) & (mag) & (mag) & (\%) & (M$_{\odot}$) \\
\hline
793 & 243.58935 & -23.65409 & 18.63 & 0.05 & 17.37 & 0.05 & 14.99 & 0.05 & 2 & 0.02/0.039 \\
924 & 244.96433 & -22.69256 & 18.45 & 0.05 & 17.03 & 0.05 & 14.42 & 0.05 & 3 & 0.027/0.056 \\
\multicolumn{11}{c}{$\cdots$}\\
1553 & 241.46574 & -22.10589 & 17.32 & 0.05 & 16.15 & 0.05 & 13.91 & 0.05 & 31 & 0.029/0.052 \\
768 & 242.76503 & -24.44526 & 19.06 & 0.05 & 17.51 & 0.05 & 14.72 & 0.05 & 8 & 0.025/0.03 \\
\hline    
\end{tabular}
\tablefoot{ID,  masses (for respectively 5 and 10Myr) and photometry from \citet{MiretRoig2022}. Full table available in electronic format.}
\end{table}

\begin{appendix}

\section{HST Photometry}

We performed aperture photometry for all targets using the \verb|photutils| Python package \citep{photutils}, adopting an aperture size of 0\farcs16 and a background annulus between 0\farcs25 and 0\farcs50. Finite-to-infinite aperture corrections provided by \citet{2022wfc..rept....2M} were applied to all measurements. Photometric uncertainties were calculated following the method outlined in \citet{Stetson1987}, accounting for Poisson noise, sky background noise, and readout noise.

For the 0\farcs11 binary J042705.86+261520.3, a larger aperture of 0\farcs44 was employed to include both components, along with a background annulus between 0\farcs44 and 0\farcs6. The corresponding finite-to-infinite aperture correction from \citet{2022wfc..rept....2M} was applied. The fluxes of the individual components were subsequently derived by measuring the flux ratio in each filter.

Finally, instrumental fluxes were converted to AB magnitudes using the formula specified in the instrument manual.:
\begin{multline}
\text{ABmag} = -2.5 \times \log_{10}(\text{Flux} \times \text{PHOTFLAM}) - 21.10  \\
- 5 \times \log_{10}(\text{PHOTPLAM}) + 18.6921 
\end{multline}
where \verb|PHOTFLAM| is the bandpass unit response  and \verb|PHOTPLAM| is the bandpass pivot wavelength (in angstroms), both provided by STSci in the FITS image headers. Figure~\ref{fig:targets_cmd_uvis} shows the (F814W, F814W-F850LP) color-magnitude diagrams for the two samples.

\section{Summary of literature surveys}

\begin{table*}
\caption{List of Taurus late type members observed at high spatial resolution.\label{tab:literature-taurus}}
\tiny
\begin{tabular}{ccccccccccc}
\hline \hline
Name & RA (J2000) & Dec (J2000) & i & $\sigma_i$ & z & $\sigma_z$ & SpType & Ref. SpType & Multiple? & Ref. Mult. \\
     & (deg) & (deg) & (mag) & (mag) & (mag) & (mag) & & & &\\
\hline
V928 Tau & 68.07858 & 24.37419 &  &  &  &  & M0.8 & (4) & Yes & (2) \\
2MASS J04414565+2301580 & 70.44022 & 23.0328 & 13.366 & 0.05 & 12.517 & 0.05 & M4.3 & (4) & Yes & (1) \\
2MASS J04390163+2336029 & 69.7568 & 23.60083 & 13.678 & 0.001 & 12.925 & 0.006 & M4.9 & (4) & No & (1),(2) \\
2MASS J04554757+3028077 & 73.94822 & 30.46881 & 13.925 & 0.003 & 13.001 & 0.009 & M4.75 & (4) & Yes & (1),(2) \\
\multicolumn{11}{c}{$\cdots$}\\
2MASS J04354526+2737130 & 68.9386 & 27.6203 & 19.757 & 0.032 & 18.043 & 0.025 & M9.25 & (1) & No & (1) \\
2MASS J04151471+2800096 & 63.81131 & 28.00267 & 19.893 & 0.016 & 18.15 & 0.014 & M8.5 & (4) & No & (1) \\
2MASS J04355143+2249119 & 68.9643 & 22.81999 & 20.428 & 0.043 & 18.697 & 0.021 & M8.5 & (4) & No & (1) \\
2MASS J04272799+2612052 & 66.86665 & 26.20146 & 20.443 & 0.04 & 18.727 & 0.026 & M9.5 & (4) & No & (1) \\
2MASS J04574903+3015195 & 74.4543 & 30.25542 & 20.798 & 0.066 & 18.97 & 0.037 & M9.25 & (4) & No & (1),(2) \\
2MASS J04215450+2652315 & 65.4771 & 26.87542 & 20.817 & 0.05 & 18.999 & 0.023 & M8.5 & (4) & No & (1),(2) \\
2MASS J04335245+2612548 & 68.46857 & 26.21524 & 20.822 & 0.031 & 19.134 & 0.019 & M8.25 & (4) & No & (1),(2) \\
2MASS J04190126+2802487 & 64.75529 & 28.04686 & 21.066 & 0.079 & 19.657 & 0.034 & M9 & (4) & No & (1) \\
2MASS J04373705+2331080 & 69.40437 & 23.51891 & 22.592 & 0.05 & 21.065 & 0.169 & L0 & (4) & No & (1) \\
\hline
\end{tabular}
\tablebib{(1)  \citet{Todorov2014}; (2) \citet{Kraus2012}; (3) \citet{Konopacky2007} ; (4) \citet{Luhman2023b}; (5) \citet{Slesnick2008}}
\tablefoot{Full table available in electronic format.}

\end{table*}

\begin{table*}
\caption{List of Usco late type members observed at high spatial resolution.\label{tab:literature-usco}}
\tiny
\begin{tabular}{ccccccccccc}
\hline \hline
Name & RA (J2000) & Dec (J2000) & i & $\sigma_i$ & z & $\sigma_z$ & SpType & Ref. SpType & Multiple? & Ref. Mult. \\
     & (deg) & (deg) & (mag) & (mag) & (mag) & (mag) & & & &\\
\hline
2MASS J15565545-2258403 & 239.23108 & -22.97788 & 12.349 & 0.05 & 12.314 & 0.05 & M0.5 & (5) & No & (3) \\
2MASS J16074449-2036030 & 241.93539 & -20.60086 & 12.732 & 0.05 & 12.12 & 0.05 & M4 & (5) & Yes & (3) \\
2MASS J16094644-1937361 & 242.44351 & -19.6267 & 12.832 & 0.05 & 12.378 & 0.05 & M1 & (5) & No & (3) \\
\multicolumn{11}{c}{$\cdots$}\\
2MASS J16081843-2232248 & 242.07682 & -22.54023 & 20.821 & 0.05 & 18.994 & 0.05 & M9.25 & (5) & No & (1),(4) \\
2MASS J16071478-2321011 & 241.81159 & -23.35031 & 21.186 & 0.05 & 19.38 & 0.05 & M9.25 & (5) & No & (1),(4) \\
2MASS J16391914-2534093 & 249.82978 & -25.56928 & 21.227 & 0.05 & 20.777 & 0.132 & M9.5 & (5) & No & (1),(4) \\
2MASS J16073799-2242468 & 241.90829 & -22.71301 & 21.422 & 0.05 & 19.651 & 0.05 & M9 & (5) & No & (1),(4) \\
2MASS J16072782-2239040 & 241.86593 & -22.65113 & 21.597 & 0.05 & 19.774 & 0.05 & M9.25 & (5) & No & (1),(4) \\
2MASS J16122895-2159358 & 243.12063 & -21.99328 & 22.026 & 0.05 & 19.891 & 0.05 & M9.5 & (5) & No & (1),(4) \\
2MASS J16130232-2124283 & 243.25968 & -21.40788 & 22.11 & 0.05 & 20.129 & 0.05 & M9.5 & (5) & No & (1),(4) \\
UGCS J160843.43-224516.0 & 242.181 & -22.75444 & 22.379 & 0.05 & 21.033 & 0.05 & L1 & (5) & No & (1),(4) \\
UGCS J160918.67-222923.8 & 242.32788 & -22.48992 & 22.924 & 0.05 & 21.008 & 0.05 & L1 & (5) & No & (1),(4) \\
\hline
\end{tabular}
\tablebib{(1)  \citet{Todorov2014}; (2) \citet{Kraus2012}; (3) \citet{Bouy2006} ; (4) \citet{Biller2011}; (5) \citet{Luhman2025} ; (6) \citet{Martin2004}}
\tablefoot{Full table available in electronic format.}

\end{table*}

\end{appendix}

\end{document}